\documentclass[pre,aps,twocolumn,showpacs]{revtex4}
\usepackage{epsfig}
\usepackage{graphicx}
\begin{document}
\title{Phase diagram of van der Waals-like phase separation in a driven granular gas}
\author{Evgeniy Khain$^{1}$, Baruch Meerson$^{1}$, and Pavel V. Sasorov$^{2}$}
\affiliation{$^{1}$Racah Institute of Physics, Hebrew University
of Jerusalem, Jerusalem 91904, Israel}
\affiliation{$^{2}$Institute of Theoretical and Experimental
Physics, Moscow 117218, Russia}
\begin{abstract}
Equations of granular hydrostatics are used to compute the phase
diagram of the recently discovered van der Waals-like phase
separation in a driven granular gas. The model two-dimensional
system consists of smooth hard disks in a rectangular box,
colliding inelastically with each other and driven by a ``thermal"
wall at zero gravity. The spinodal line and the critical point of
the phase separation are determined. Close to the critical point
the spinodal and binodal (coexistence) lines are determined
analytically. Effects of finite size of the confining box in the
direction parallel to the thermal wall are investigated. These
include suppression of the phase separation by heat conduction in
the lateral direction and a change from supercritical to
subcritical bifurcation.

\end{abstract}
 \maketitle

\section{Introduction}

Granular gases (gases of inelastically colliding macroscopic
particles) exhibit a plethora of symmetry-breaking instabilities
and clustering phenomena \cite{reviews}. Their investigation is
useful both for testing and improving the models of granular flow,
and for a deeper understanding of far-from-equilibrium dynamics in
general. In this work we focus our attention on a recently
discovered phase separation instability which occurs in a
prototypical two-dimensional granular system: an assembly of
monodisperse hard disks in a box, colliding inelastically with
each other and driven, at zero gravity, by a rapidly vibrating or
thermal wall. An immediate consequence of the inelasticity of the
particle collisions is the formation of a laterally uniform
cluster (the stripe state) at the wall opposite to the driving
wall \cite{Grossman,Kudrolli1}. Both granular hydrodynamics and
direct molecular dynamic simulations show that this simple
clustering state can exhibit spontaneous symmetry-breaking
instability leading to phase separation: coexistence of dense and
dilute regions of the granulate (droplets and bubbles) along the
wall opposite to the driving wall
\cite{LMS,Brey2,KM,Argentina1,LMS2,MPSS,Argentina2}. This
far-from-equilibrium phase separation is strikingly similar to the
gas-liquid transition in the classical van der Waals model. The
objective of this work is a systematic investigation of the steady
states in this system and computation of the phase diagram
starting from the Navier-Stokes granular hydrodynamics. Granular
hydrodynamics is expected to be an accurate leading order theory
when the mean free path of the particles is much less than a
characteristic inhomogeneity scale of the problem, and the mean
time between two consecutive collisions of a particle is much less
than any time scale the hydrodynamics attempts to describe. In
addition, we should work with sufficiently low particle densities,
when an account of binary collisions and volume exclusion effects
is sufficient \cite{Jenkins}. As will be shown below (see also
Ref. \cite{Grossman}), the requirement that the mean free path be
small compared to the inhomogeneity scale implies in this system
that particle collisions must be nearly elastic.

For steady states with a zero mean flow, granular hydrodynamic
equations reduce to granular \textit{hydrostatics}, see Section 2.
The hydrostatic problem is fully described by three scaled
parameters introduced below: the area fraction of the particles
$f$, the aspect ratio of the box $\Delta$ and the effective
hydrodynamic inelastic heat loss parameter $\Lambda$
\cite{LMS,KM}. By solving the hydrostatic equations numerically,
we obtain, in Section 2, the spinodal line and the critical point
of the phase separation in the limit of $\Delta \to \infty$.
Section 3 deals with the same limit of $\Delta \to \infty$ but, in
addition, assumes a close proximity to the critical point. Here we
find the spinodal and binodal (coexistence) lines analytically and
determine the structure of the steady-state domain wall,
separating the more dense and less dense stripes in the lateral
direction. Effects of finite aspect ratio $\Delta$ are addressed
in Sec. 4. These include suppression of the phase separation by
heat conduction in the lateral direction and a change from a
supercritical to subcritical bifurcation. Section 5 includes a
brief discussion of the results and proposes directions of future
work.

\section{Hydrostatics of phase separation: model and general results}

In this Section we formulate the model, briefly review the stripe
state and compute the spinodal line and the critical point of the
phase separation for a laterally infinite system, $\Delta \to
\infty$.

\subsection{Model}

Consider an assembly of inelastically colliding hard disks of
diameter $d$ and mass $m=1$, moving in a box with dimensions $L_x
\times L_y$ at zero gravity. Collisions of disks with the walls
$x=0$ and $y=\pm\, L_y/2$ are assumed elastic. Alternatively,
periodic boundary conditions in the $y$-direction can be imposed.
We refer to the $y$-direction as the lateral direction of the
system. In a laterally infinite system $L_y \to \infty$. The
particles are driven by a ``thermal" wall: a constant granular
temperature $T_0$ is prescribed at $x=L_x$. The inelasticity of
the particle collisions is parametrized by a constant normal
restitution coefficient $r$; we will work in the nearly elastic
limit $1-r^2 \ll 1$. We also assume a moderate number density $n$:
$n/n_c < 0.5$, where $n_c = 2/(\sqrt{3} d^2)$ is the hexagonal
close packing density. The last two assumptions allow us to employ
the Navier-Stokes granular hydrodynamics. Throughout this work, we
will use for concrete calculations the constitutive relations
suggested by Jenkins and Richman \cite{Jenkins}. These relations
were derived by analogy with those obtained in the framework of a
successful but still empiric Enskog theory \cite{Resibois}.
Detailed comparisons with molecular dynamic simulations shows that
the error margin of the Enskog heat conductivity can reach as much
as 10 - 15 percent \cite{KMSunpublished}. Still, these relations
seem to be the best available constitutive relations for moderate
densities. Importantly, most of the analytical results in this
work are written in a more general form which only assumes a
Navier-Stokes structure of hydrodynamic equations. The
Jenkins-Richman's relations are used only for computing numerical
factors.

Energy input at the thermal wall balances the energy loss due to
inter-particle collisions, so we assume that the system reaches a
steady state with a zero mean flow. Therefore, the full
hydrodynamic equations reduce to hydrostatic equations:
\begin{equation}
p=const\;\;\; \mbox{and} \;\;\;\nabla\cdot(\kappa\nabla T)=I\,,
\label{E1}
\end{equation}
where $p$ is the granular pressure, $T$ is the granular
temperature, $\kappa$ is the thermal conductivity, and $I$ is the
rate of energy loss by collisions. Notice that we did not account,
in the second of  Eq. (\ref{E1}), for an additional (inelastic)
contribution to the heat flux which is proportional to the
\textit{density} gradient.  For a dilute gas this term was derived
in Ref. \cite{BDKS}. It can be neglected in the nearly elastic
limit which is assumed throughout this paper.

The constitutive relations entering Eqs. (\ref{E1}) include the
equation of state $p = p (n, T )$ and expressions for $\kappa$ and
$I$ in terms of $n$ and $T$. In our notation, these relations can
be written as \cite{Jenkins}
\begin{eqnarray}
p&=&nT(1+2G^{\prime})\,,\nonumber
\\
\kappa&=&\frac{2dnT^{1/2}G^{\prime}}{\sqrt{\pi}}
\left[1+\frac{9\pi}{16}\left(1+\frac{2}{3G^{\prime}}
\right)^2\right]\,,\nonumber
\\
I&=&\frac{8(1-r)nT^{3/2}G^{\prime}}{\sqrt{\pi}d}\,,\nonumber
\\
G^{\prime}&=&\frac{\nu\left(1-\frac{7\nu}{16}\right)}{(1-\nu)^2},
\label{babic}
\end{eqnarray}
where $\nu=n(\pi d^2/4)$ is the solid fraction. Let us rescale the
coordinates by $L_x$: ${\mathbf{r}}/L_x \to \mathbf{r}$. In the
rescaled coordinates the box dimensions become $1 \times \Delta$,
where $\Delta=L_y/L_x$ is the aspect ratio of the box. Introducing
a normalized inverse density $z(x,y)=n_c/n(x,y)$ and eliminating
the temperature, one can rewrite Eqs. (\ref{E1}) as a single
equation for $z(x,y)$ \cite{Grossman,LMS,KM}:
\begin{equation}
\nabla\cdot(F(z)\,\nabla z)=\Lambda \, Q(z), \label{E4}
\end{equation}
where $F(z)= A(z)\,B(z)$,
\begin{eqnarray}
A(z)&=&\frac{G\left[1+\frac{9\pi}{16}
\left(1+\frac{2}{3G}\right)^2\right]}{z^{1/2}(1+2G)^{5/2}}\,,\nonumber
\\
B(z)&=&1+2G+\frac{\pi}{\sqrt{3}}
\frac{z(z+\frac{\pi}{16\sqrt{3}})}{(z-\frac{\pi}{2\sqrt{3}})^3}\,,\nonumber
\\
Q(z)&=&\frac{6}{\pi}\frac{z^{1/2}G}{(1+2G)^{3/2}}\,,\nonumber
\\
G(z)&=&\frac{\pi}{2\sqrt{3}}
\frac{z-\frac{7\pi}{32\sqrt{3}}}{(z-\frac{\pi}{2\sqrt{3}})^2}\,.
\label{E5}
\end{eqnarray}
The parameter
\begin{equation}
\Lambda=\frac{2\pi}{3}(1-r)\left(\frac{L_x}{d}\right)^2\,,
\label{E5a}
\end{equation}
which appears in the right hand side of Eq. (\ref{E4}), is the
hydrodynamic inelastic heat loss parameter. Notice that it can be
made arbitrary large (by taking large enough $L_x/d$), no matter
how small the inelasticity $q=(1-r)/2$ is.

Now we specify the boundary conditions for Eq. (\ref{E4}). At the
elastic walls $x = 0$ and  $y = \pm \,\Delta/2$ the normal
component of the heat flux must vanish. In terms of the inverse
density $z$ one has $\nabla_n z = 0$ at these three walls. Here
index $n$ denotes the gradient component normal to the wall.
Alternatively, for the periodic boundary conditions we should
demand $z(x,y+2 \pi/\Delta)=z(x,y)$.  The constant temperature at
the thermal wall $x=1$ yields the simple condition
\begin{equation}\label{thermal_wall}
z(x=1,y)= const
\end{equation}
with an \textit{a priori} unknown constant. As the total number of
particles $N$ is fixed, the normalization condition
\begin{equation}
\frac{1}{\Delta}\int_0^1\int_{-\Delta/2}^{\Delta/2} \frac{dx
dy}{z(x,y)}=f \label{E6}
\end{equation}
should be imposed, where $f=\langle n \rangle / n_c$ is the area
fraction of the grains and $\langle n \rangle = N/(L_x L_y)$ is
the average number density of the grains.

Equation (\ref{E4}) with the boundary conditions at the four walls
and Eq. (\ref{E6}) make a complete set. Notice that the
steady-state {\it density} distribution is independent of the wall
temperature $T_0$ \cite{LMS}. The wall temperature only sets the
scale of the \textit{temperature} profile in the system, and
affects the steady-state \textit{pressure}. The governing
parameters of the system are the scaled numbers $\Lambda$, $f$ and
$\Delta$. For a laterally infinite system, $\Delta \to \infty$,
only two governing parameters $\Lambda$ and $f$ remain in the
hydrostatic formulation.

Let us define the scaled temperature $\tilde{T}=T/T_0$ and
pressure
\begin{equation}
\tilde{p}=\frac{p}{n_cT_0}=\tilde{T} \,\Pi(z)\, , \label{Pr10}
\end{equation}
where
\begin{equation}
\Pi(z)=\frac{1+2G(z)}{z}\,. \label{Pr20}
\end{equation}
Once the steady state density profile and the (uniform) steady
state pressure $\tilde{p}$ are found, Eq.~(\ref{Pr10}) determines
the steady state temperature profile $\tilde{T} (x,y)$.

\begin{figure}
\includegraphics[width=7.0cm,clip=]{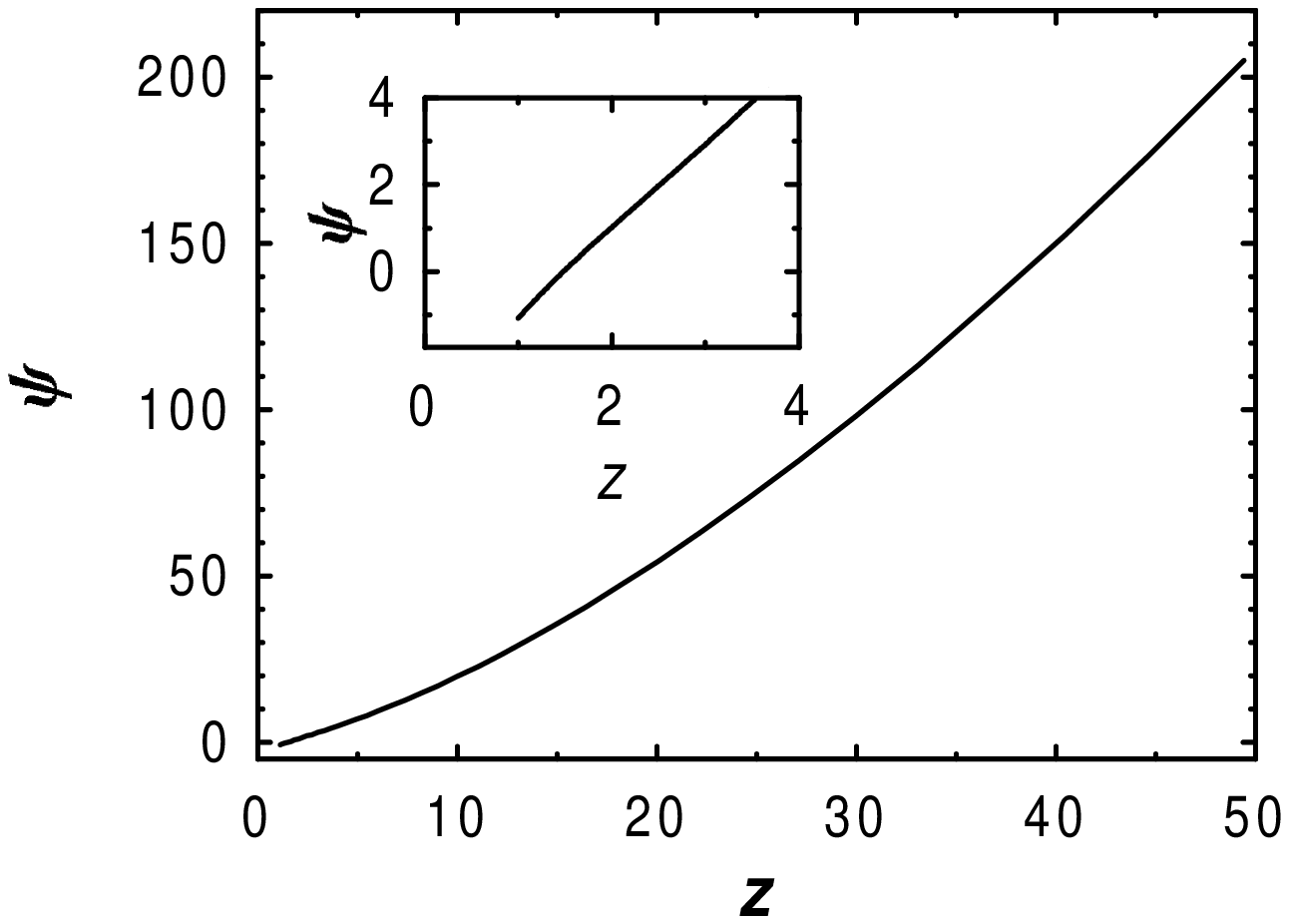}
\caption{The effective heat flux potential $\psi$ versus the
inverse scaled density $z$. The inset shows a blowup of the region
of $1<z<4$.} \label{psi}
\end{figure}

\begin{figure}
\includegraphics[width=7.0cm,clip=]{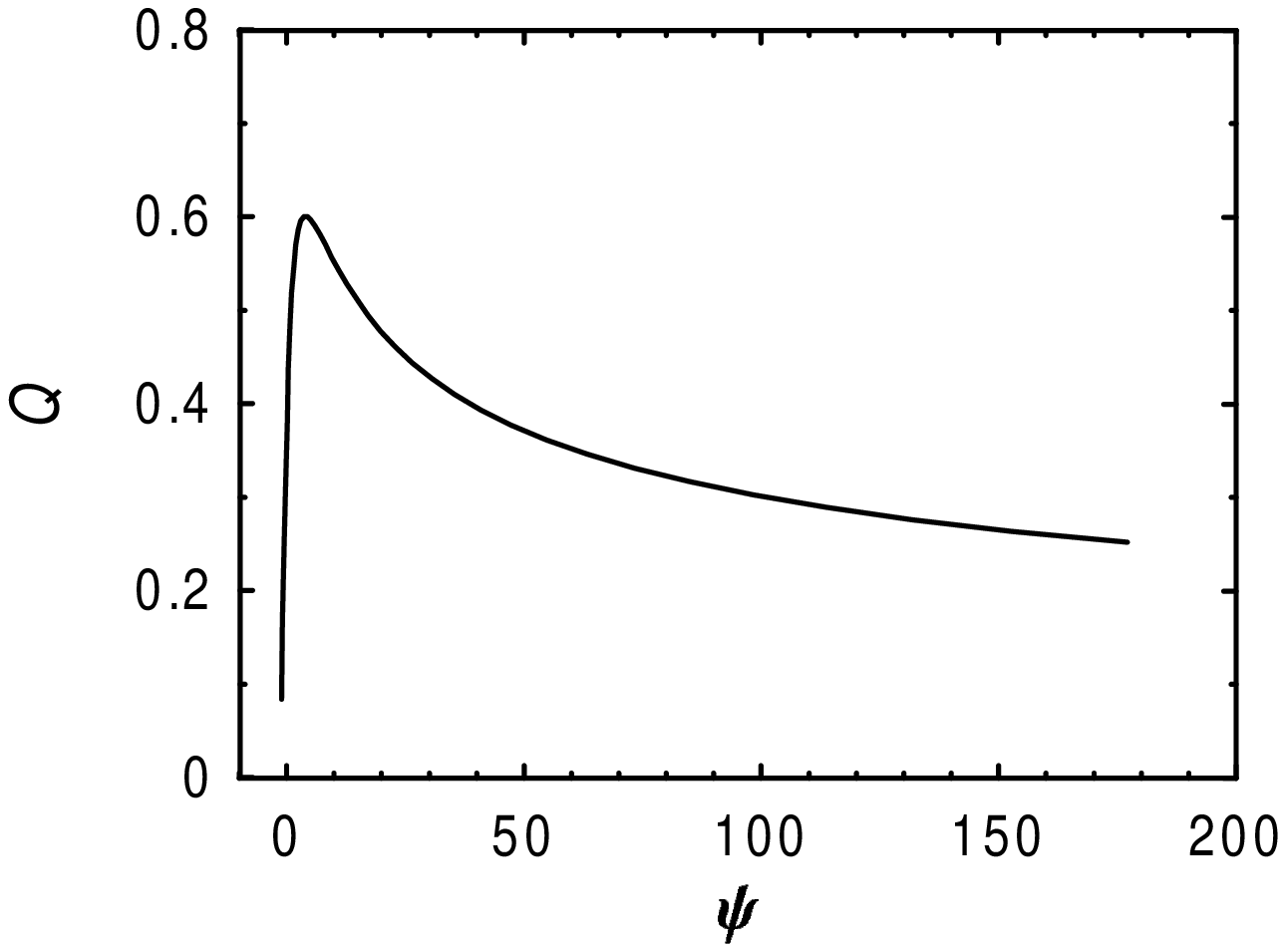}
\caption{The effective heat loss function $Q(\psi)$ which appears
on the right hand side of Eq.~(\ref{i10}).} \label{Q_psi}
\end{figure}

The vector field $F(z) \nabla z$ entering Eq. (\ref{E4}) is, up to
a sign, the scaled heat flux. Equation (\ref{E4}) gets simpler if
we introduce, as a new variable, the scalar potential of the heat
flux: $\psi=\int^{z} F(z^{\prime})\, dz^{\prime}$. To avoid a
divergence of the integral at infinity, we account for the
diverging part directly, by extracting the first two terms of
expansion of $F(z)$ at $z \to \infty$:
\begin{equation}
F(z)= \frac{\sqrt{3}}{2}z^{1/2}-\frac{7\pi}{64}\,z^{-1/2}+f(z)\,.
\label{i70}
\end{equation}
The integral of $f(z)$ already converges at infinity, and we
obtain
\begin{equation}
\psi(z)=\frac{1}{\sqrt{3}}z^{3/2}+\frac{7\pi}{32}z^{1/2}
-\int\limits_z^\infty f(z)\, dz\, , \label{i60}
\end{equation}
The potential $\psi$ grows monotonically with $z$, see
Fig.~\ref{psi}. Now Eq. (\ref{E4}) becomes a nonlinear Poisson
equation \cite{MPSS}:
\begin{equation}
\nabla^2\psi=\Lambda Q(\psi)\,, \label{i10}
\end{equation}
where by $Q(\psi)$ we actually mean $Q\left[z(\psi)\right]$ here
and in the following. The function $Q(\psi)$ is depicted in
Fig.~\ref{Q_psi}.  We will be dealing with Eq. (\ref{i10})
throughout the paper. The boundary conditions for $\psi(x,y)$ are
identical to the boundary conditions for $z(x,y)$:
\begin{eqnarray}
&\left.\frac{\partial \psi(x,y)}{\partial x}\right|_{x=0} &=0\, ,
\label{ii20}\\
&\psi(1,y)& =const\, , \label{ii30}
\end{eqnarray}
supplemented by either no-flux, or periodic boundary condition at
the walls $y= \pm \,\Delta/2$. Note that $z$ is assumed to be
expressed through $\psi$ in Eq.~(\ref{E6}).

Importantly, $Q(\psi)$ decreases with an increase of $\psi$ at
large enough $\psi$. (In the dilute limit, $\psi \gg 1$, one has
$Q(\psi) \propto \psi^{-1/3}$.) This implies non-uniqueness of
steady state solutions of Eq.~(\ref{i10}) \cite{non-unique}, and
opens the way to phase separation and coexistence.

\subsection{Stripe states, spinodal line and critical point}

\begin{figure}
\includegraphics[width=7.0cm,clip=]{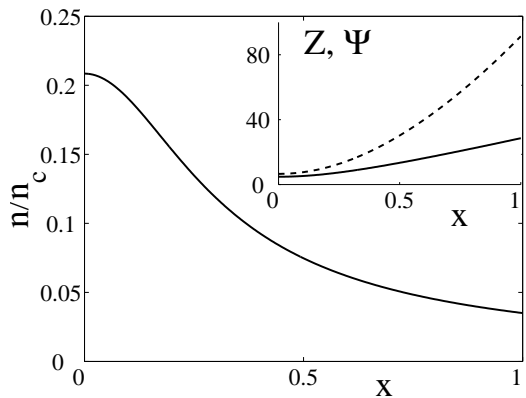}
\caption{Spatial profiles of the stripe state for $\Lambda =
344.2$ and $f = 0.095$, which are the critical values of $\Lambda$
and $f$ for the constitutive relations (\ref{E5}). Shown in the
main part of the figure is the scaled density $n(x)/n_c =
Z^{-1}(x)$. The inset shows $Z(x)$ (the solid line) and the heat
flux potential $\Psi(x)$ (the dashed line), obtained by solving
Eqs.~(\ref{cr10})-(\ref{cr21}). } \label{fig1}
\end{figure}

The simplest steady state of the system is the ``stripe": a
laterally uniform state which corresponds to the $y$-independent
solution $z = Z(x)$ \cite{Grossman}, or $\psi=\Psi(x)$. It is
described by the equation
\begin{equation}
\Psi^{\prime\prime}=\Lambda Q(\Psi), \label{cr10}
\end{equation}
with the boundary condition
\begin{equation}
\Psi^{\prime}(0)=0\, \label{cr20}
\end{equation}
and normalization condition
\begin{equation}
\int_0^1 \frac{dx}{Z[\Psi(x)]}=f\,. \label{cr21}
\end{equation}
Here and in the following the primes denote the $x$-derivatives.
As Eq. (\ref{cr10}) does not include the first derivative
$\Psi^{\prime}(x)$, it has ``energy" integral and therefore is
integrable. A numerical solution however, is more practical.
Figure ~\ref{fig1} gives an example of the density profile of the
stripe state in terms of the scaled density $n(x)/n_c = Z^{-1}(x)$
and the auxiliary functions $\Psi(x)$ and $Z(x)$ for $\Lambda =
344.2$ and $f = 0.095$, corresponding to the critical point of the
phase separation (see below).

The stripe state problem~(\ref{cr10})-(\ref{cr21}) can be recast
into a more convenient initial value problem if we use, instead of
the normalization condition (\ref{cr21}), a boundary condition
\begin{equation} \Psi(0)=a \,.\label{cr30}
\end{equation}
Indeed, the initial value problem defined by Eqs. (\ref{cr10}),
(\ref{cr20}) and (\ref{cr30}) has a unique solution
$\Psi(x,a,\Lambda)$. Having found it,  one can calculate the area
fraction $f=f(a,\Lambda)$ from Eq. (\ref{cr21}). Importantly, for
the Enskog-type constitutive relations (\ref{E5}), $f(a,\Lambda)$
turns out to be a monotonic function of $a$ for any fixed
$\Lambda$. This enables one to use the pair of numbers $(a,
\Lambda)$ instead of $(f,\Lambda)$ for the parametrization of all
possible stripe states. The parametrization $(a, \Lambda)$ will be
often used in the following.

\begin{figure}
\includegraphics[width=7.0cm,clip=]{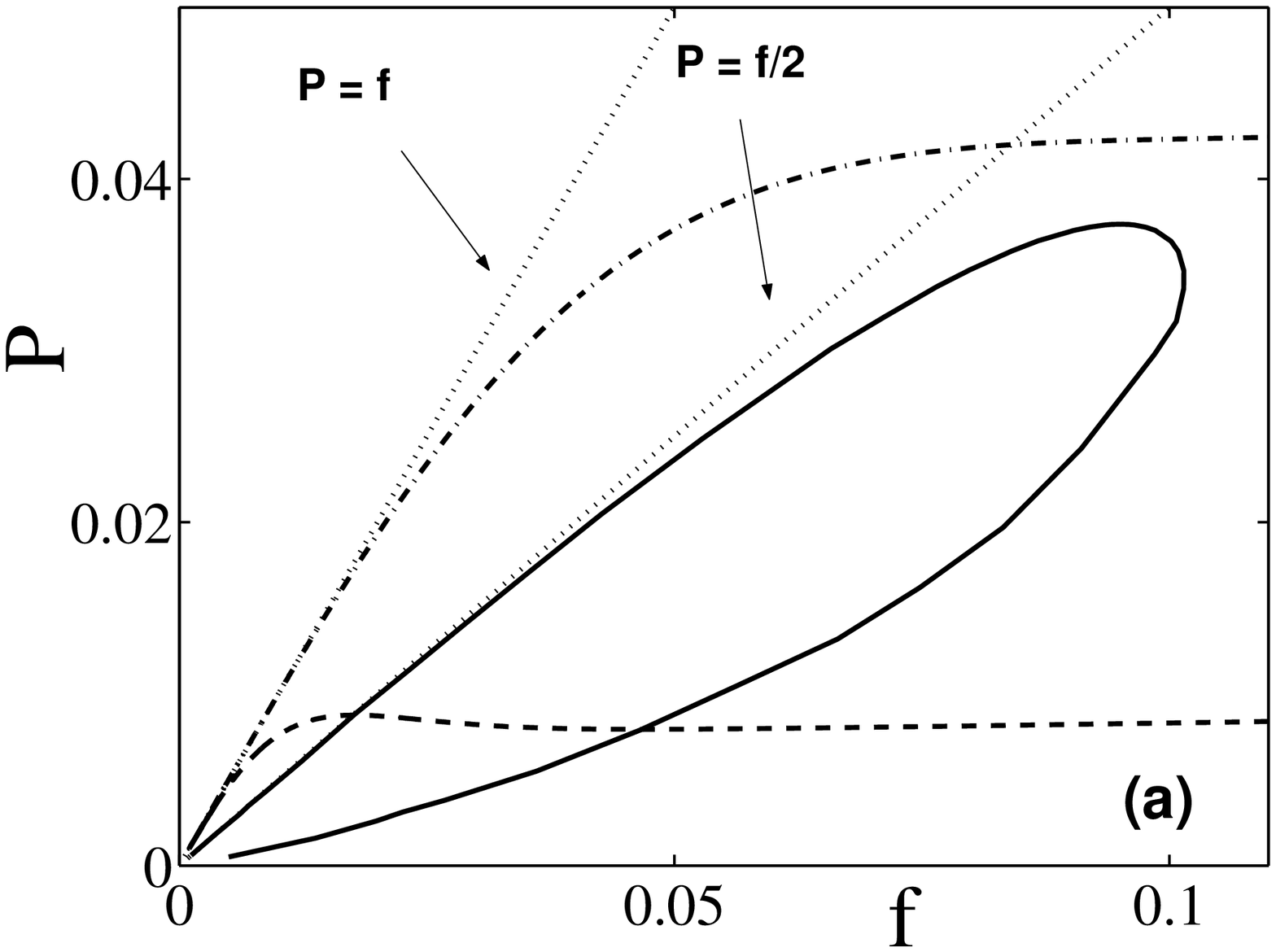}
\includegraphics[width=7.0cm,clip=]{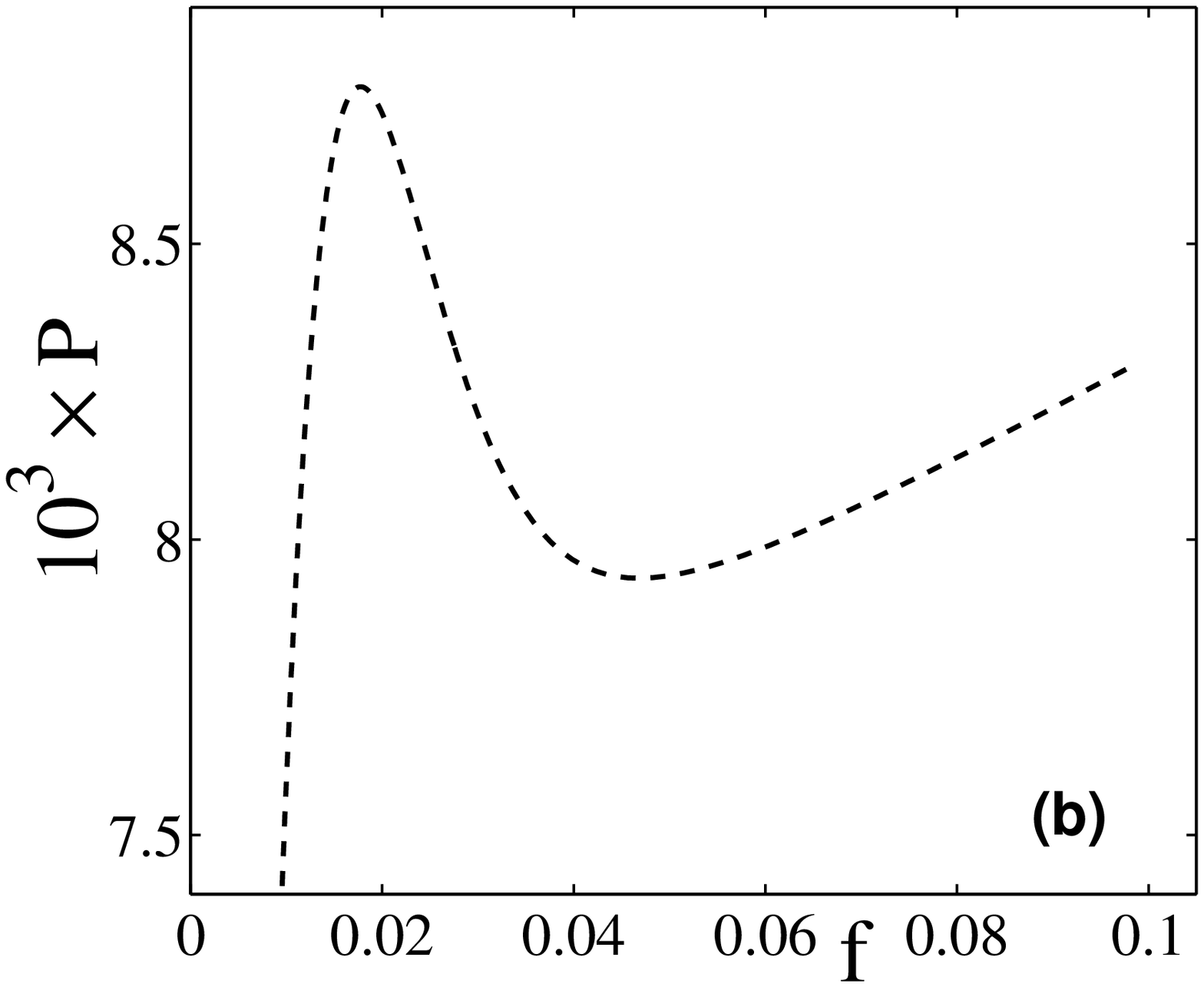}
\caption{The spinodal line (the solid line) and the $P(f)$ curves
for the stripe state, for $\Lambda = 5 \cdot 10^3$ and $\Lambda =
280$ (the dashed and dash-dotted lines, respectively) (a). The two
dotted straight lines are the dilute-limit asymptotes
$P_1(\Lambda, f) = f$ and $P_2(\Lambda, f) = f/2$ (a). Figure b
shows the $P(f)$ curve for $\Lambda = 5 \cdot 10^3$ in a different
scale, to make the region of negative compressibility $dP/df<0$
more visible.}\label{diagram1}
\end{figure}

\begin{figure}
\includegraphics[width=7.0cm,clip=]{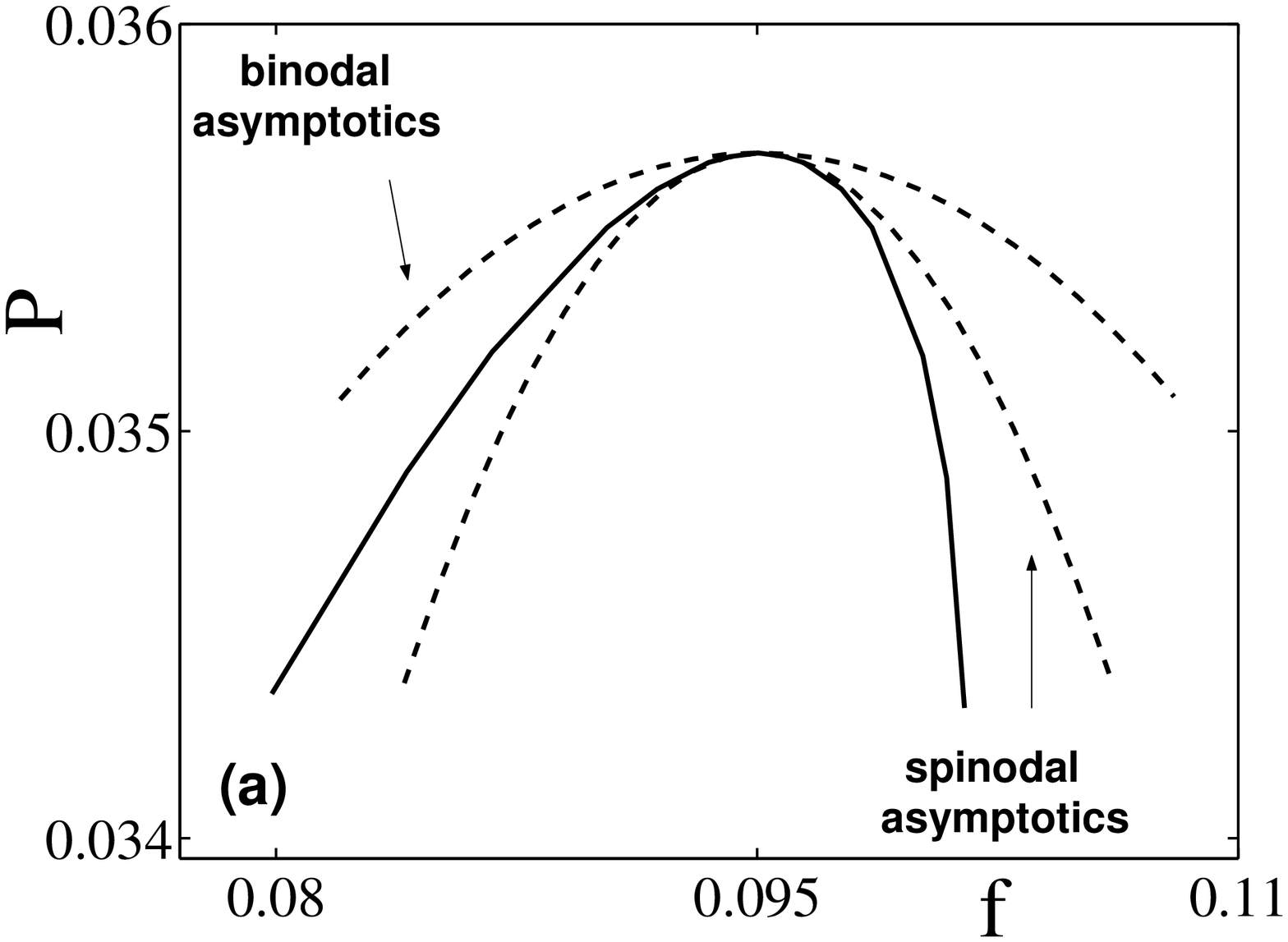}
\includegraphics[width=7.0cm,clip=]{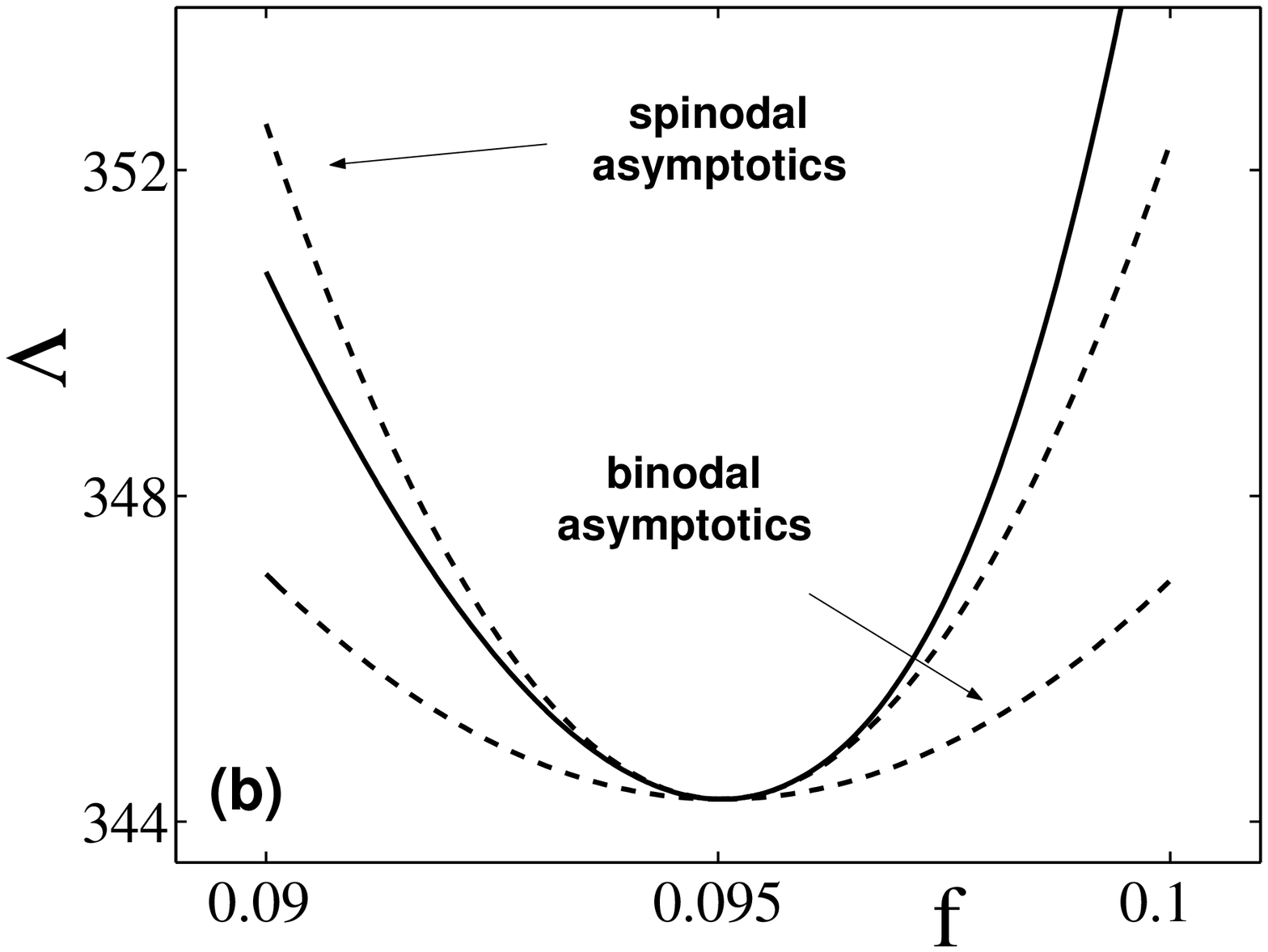}
\caption{The spinodal line (the solid line) and the spinodal and
binodal asymptotics close to the critical point (the dashed lines)
in the variables $f, P$ (a) and $f, \Lambda$ (b).}
\label{diagram2}
\end{figure}

In a wide region of the parameter space $(f,\Lambda,\Delta)$ the
stripe state undergoes a phase separation instability  and gives
way to a laterally asymmetric state
\cite{LMS,Brey2,Argentina1,KM,LMS2,MPSS,Argentina2}. The
instability is driven by negative compressibility of the stripe
state in the lateral direction ~\cite{Argentina1,KM}, resulting
from energy loss in particle collisions. Let $P=\tilde{T}
\,\Pi(Z)$ be the scaled steady-state gas pressure of the stripe
state. In the limit of $\Delta \to \infty$, the spinodal region in
the $(f,\Lambda)$-plane is defined by the condition
$\left(\partial P/\partial f\right)_{\Lambda}<0$. As $P$ is
constant in space, it can be conveniently computed at the thermal
wall $x=1$ \cite{KM}. Here $\tilde{T} =1$ and therefore
\begin{equation}
P=P(a,\Lambda)=\left.\Pi[z(\psi)]\right|_{\psi=\Psi(1,a,
\Lambda)\,.}  \label{Pr20a}
\end{equation}
Alternatively,
\begin{equation}
P (f,\Lambda) = \frac{1+ 2 G[Z_1(f,\Lambda)]}{Z_1(f,\Lambda)}\,,
\label{stripe_pressure}
\end{equation}
where $Z_1 = Z(x=1)$. The spinodal line of the phase separation
(again, in the limit of $\Delta \to \infty$) is determined by the
condition
\begin{equation}
\frac{\partial P(f,\Lambda)}{\partial f}=0\,. \label{cr31}
\end{equation}
Solving Eqs. (\ref{cr10})-(\ref{cr30}), we computed from Eq.
(\ref{stripe_pressure}) the $P(f)$ curves at different $\Lambda$,
see Fig.~\ref{diagram1}a. These computations yield a critical
point $(f_c,P_c)$ [equivalently, $(f_c,\Lambda_c)$, or
$(a_c,\Lambda_c)$].  For $\Lambda < \Lambda_c$ the pressure $P$
increases monotonically with $f$ (like in an elastic gas), so
there is no phase separation instability. For $\Lambda
> \Lambda_c$ the pressure $P$ versus $f$ has a non-monotonic part.
The maximum and minimum points of $P(f)$ at different $\Lambda$
yield the spinodal line. This line in the $(f,P)$-plane is shown
by the solid curve in Fig.~\ref{diagram1}a. Figure~\ref{diagram1}a
also shows two straight lines. These are the dilute-limit
asymptotes of $P(f)$ and of the low-density branch of the spinodal
line, respectively (see the next subsection).
Figure~\ref{diagram1}b shows, on a different scale, the
$P(f)$-dependence for a fixed $\Lambda$ within the spinodal
region, to make the region of negative lateral compressibility
more visible. Figure ~\ref{diagram2} depicts a part of the
spinodal line, together with the asymptotics of the spinodal and
binodal (coexistence) lines in the vicinity of the critical point,
found analytically in Section 3.

Now let us again use the $a$-parametrization of the stripe
solution. Equation (\ref{cr31}) for the spinodal line is
equivalent to
\begin{equation}\frac{\partial P(a,\Lambda)}{\partial a}=0\, .
\label{cr32}
\end{equation}
The critical point is the merging point of the maximum and minimum
points of the $P(f)$ [or $P(a)$] curve at different $\Lambda$.
Therefore, the critical point is defined by the following two
conditions:
\begin{eqnarray}
\frac{\partial P(a,\Lambda)}{\partial a}(a_c,\Lambda_c)&=&0\, ,
\label{cr34}\\
\frac{\partial^2 P(a,\Lambda)}{\partial a^2}(a_c,\Lambda_c)&=&0\, .
\label{cr36}
\end{eqnarray}
For an assembly of hard spheres below the freezing point one has
$\left(\partial p/\partial n\right)_T>0$, which follows $(dP/dZ)|_
{Z=Z_1(a,\Lambda)} \neq 0$. Therefore, Eq.~(\ref{cr32}) for the
spinodal line is equivalent to
\begin{equation}\frac{\partial \Psi (1,a,\Lambda)}{\partial
a}=0\,,
\label{cr32a}
\end{equation}
while Eqs.~(\ref{cr34}) and (\ref{cr36}) for the critical point
can be rewritten as
\begin{eqnarray}
\frac{\partial \Psi}{\partial a}(1,a_c,\Lambda_c)&=&0\, ,
\label{cr40}\\
\frac{\partial^2 \Psi}{\partial a^2}(1,a_c,\Lambda_c)&=&0
\label{cr50}
\end{eqnarray}
(recall that the first argument of function $\Psi$ and of its
derivatives stands for $x$). Solving Eqs.~(\ref{cr10}),
(\ref{cr20}), (\ref{cr30}), (\ref{cr40}) and~(\ref{cr50})
numerically, we find the critical point
\begin{eqnarray}
a_c&=&6.580\dots\, ,
\label{n30}\\
\Lambda_c&=&344.2\dots\,. \label{n20}
\end{eqnarray}
Using Eqs.~(\ref{cr21}) and (\ref{stripe_pressure}), we find the
critical point in the variables $f,P$:
\begin{eqnarray}
f_c&=& f(a_c,\Lambda_c)= 0.0950 \dots\, ,
\label{n35}\\
P_c&=&P(a_c,\Lambda_c) = 0.0373 \dots\, . \label{n38}
\end{eqnarray}
We checked that the maximum density of the stripe states,
corresponding to the spinodal shown in Fig.~\ref{diagram1}a, is
less than $0.5\,n_c$, that is within the assumed validity domain
of the constitutive relations (\ref{E5}). At the critical point
itself the maximum scaled density of the stripe state is
$n(x=0,a_c,\Lambda_c)/n_c=1/Z(x=0,a_c,\Lambda_c)= 0.2086\dots$, a
moderate value.

The critical point, predicted by the Enskog-type granular
hydrodynamics, agrees fairly well with molecular dynamic
simulations by Soto \textit{et al.} \cite{Argentina1,Argentina2}.
The issues of accuracy of the hydrodynamic results are discussed
in Section 5.

\subsection{Dilute-gas limit}
In the dilute-gas limit, $Z \gg 1$, we can obtain  the two
straight-line asymptotes shown in Fig.~\ref{diagram1}a: of the
stripe pressure $P(f,\Lambda)$ and of the low-density branch of
the spinodal line. Here the stripe state equation is
\begin{equation}
\frac{d^2}{d \bar{x}^2} Z^{3/2} = 3 Z^{-1/2}\,, \label{E10}
\end{equation}
where a different rescaling of the coordinate is introduced:
$\bar{x} = x \sqrt{\Lambda}$. (Notice that in the physical
coordinate $x_{phys}$ the new rescaling does not include $L_x$.)
The boundary conditions are
$$
Z(\bar{x}=0)=Z_0 \;\;\;\;\;\;\mbox{and} \;\;\;\;\;\;
\frac{dZ}{d\bar{x}}(\bar{x}=0) = 0\,,
$$
where $Z_0$ is related to $f$ and $\Lambda$ by the normalization
condition $\int^{\sqrt{\Lambda}}_0 Z^{-1} d\bar{x} = f
\Lambda^{1/2}$. The solution to this problem is \cite{KM}
\begin{equation}
\bar{x}=\frac{Z_0}{2}\left(\mbox{arccosh}\sqrt{\zeta}+\sqrt{\zeta^2-\zeta}\right)\,,
\label{E11}
\end{equation}
where $\zeta=Z/Z_0$ and
\begin{equation}
Z_0= \frac{4\Lambda^{1/2}}{2 f\Lambda^{1/2}+\sinh
(2f\Lambda^{1/2})}\,. \label{Z0}
\end{equation}
In the dimensional units the density profile (\ref{E11}) is
determined by a single parameter: $\xi=f \Lambda^{1/2}$. The
scaled pressure of the stripe state is
\begin{equation}
\label{dilutepressure}
 P= \frac{1}{Z_1} =
\frac{2\xi+ \sinh (2 \xi)}{4 \Lambda^{1/2} \cosh^2 \xi},
\end{equation}
where $Z_1 = Z(x=1) = Z_0 \cosh^2 \xi$. Assuming $\xi \ll 1$, we
obtain from  Eq. (\ref{dilutepressure}) the dilute-limit asymptote
$P(f,\Lambda) = f$.

The low-density asymptote of the spinodal is determined from the
condition
$$
\left(\frac{\partial P}{\partial f}\right)_{\Lambda} =
\Lambda^{1/2} \left(\frac{\partial P}{\partial\xi}
\right)_{\Lambda} = 0\,.
$$
This equation, together with Eq. (\ref{dilutepressure}), yields
$\coth(\xi) = \xi$ \cite{KM}. Substituting this result back into
Eq. (\ref{dilutepressure}), we obtain $P(f, \Lambda) = f/2$.

Using the dilute-limit Eqs. (\ref{E11}) and (\ref{Z0}), one can
verify that validity of granular hydrodynamics as an accurate
leading order theory in this problem indeed demands nearly elastic
collisions (see also Ref. \cite{Grossman}). Equation (\ref{E11})
implies that the inverse scaled density $Z(x)$ can be presented as
$Z=Z_0\, {\cal F} (\bar{x}/Z_0)$, where function ${\cal F}$ does
not include any additional parameters. Therefore, the
characteristic scale of inhomogeneity of the stripe state in the
$x$-direction in the rescaled coordinate $\bar{x}$ is of the order
of $Z_0$. Going back to the physical coordinate,
$\bar{x}={\Lambda}^{1/2} x_{phys}/L_x \sim q^{1/2} x_{phys}/d$, we
see that the inhomogeneity scale is of the order of $Z_0 d/q^{1/2}
\sim (n_0 d \,q^{1/2})^{-1}$. For hydrodynamics to be valid, this
quantity should be much greater than the characteristic mean free
path of the particles $\bar{l}$ which is of the order of $(n_0
d)^{-1}$. Therefore, the validity of hydrodynamics in this system
requires $q^{1/2} \ll 1$, quite a stringent condition. An
additional discussion of this condition is presented in Sec. V.

\section{Spinodal and binodal lines in the vicinity of the
critical point}

\subsection{Mathematical preliminaries}

In the vicinity of the critical point $a_c, \Lambda_c$ (at
$\Lambda>\Lambda_c$) the two-dimensional, phase separated state is
very close to the one-dimensional stripe state. Therefore, the
spindal and binodal lines can be obtained by expanding $\psi(x,y)$
around the stripe-state solution $\Psi(x,a,\Lambda)$ in the power
series of $a-a_c$ and $\Lambda-\Lambda_c$. For brevity of notation
the subscripts $a$ and $\Lambda$ will denote the partial
derivatives $\partial/\partial a$ and $\partial/\partial \Lambda$,
respectively, while the prime $^{\prime}$ will stand for the
partial derivative $\partial/\partial x$ as before. As will become
clear shortly, the following functions of $x$ at the critical
point need to be computed:
$$\Psi^{\prime},\quad \Psi_a,\quad \Psi_{\Lambda},\quad \Psi_{a\Lambda},
\quad \Psi_{aa} \quad \mbox{and} \quad \Psi_{aaa}\,.$$ As the
stripe solution $\Psi(x,a,\Lambda)$ is available in quadrature,
the derivatives of $\Psi$ with respect to parameters $a$ and
$\Lambda$ are known. More practical, however, is a different
approach. As shown in Appendix, all these functions [and an
additional function $\Phi(x)$ that we will need, see Eq.
(\ref{bin110}) below] can be expressed as solutions of the linear
differential equation
\begin{equation}
\label{gen_form}
 w^{\prime\prime}(x)-\Lambda_c\, Q_\psi\left[\Psi(x,a_c,\Lambda_c)\right]\ w(x) = S(x)
\end{equation}
with different source terms $S(x)$ and different boundary
conditions. For the first derivatives, $\Psi^{\prime}$ and
$\Psi_a$, the source term vanishes. Therefore, the rest of the
functions can be expressed through $\Psi^{\prime}$ and $\Psi_a$,
see Appendix.

\subsection{Spinodal line in the vicinity of the critical point}
\label{spcr}

It follows from Eqs.~(\ref{cr40}) and~(\ref{cr50})  that the first
non-vanishing term in the expansion of $\Psi(x,a,\Lambda)$ in the
powers of $a-a_c$ at the critical point is the cubic term
$(a-a_c)^3$. Therefore, we should keep the following terms in the
expansion:
$$
\Psi(x,a,\Lambda)=\Psi(x,a_c,\Lambda_c)
+\Lambda_c\, \Psi_\Lambda(x,a_c,\Lambda_c)\, \delta
$$
$$
+ a_c\, \Psi_a(x,a_c,\Lambda_c)\, u +
\frac{a_c^2}{2}\,\Psi_{aa}(x,a_c,\Lambda_c)\,u^2
$$
\begin{equation}
+a_c\Lambda_c\,\Psi_{a\Lambda}(x,a_c,\Lambda_c)\,u\delta
+\frac{a_c^3}{6}\,\Psi_{aaa}(x,a_c,\Lambda_c)\,u^3\,, \label{cr52}
\end{equation}
where an order parameter $u=a/a_c-1$ and control parameter
$\delta=\Lambda/\Lambda_c-1$ have been introduced.

To obtain the equation of the spinodal line we differentiate
Eq.~(\ref{cr52}) with respect to $a$, put $x=1$ and use
Eqs.~(\ref{cr32a})-(\ref{cr50}). The result is
\begin{equation}
\delta=-\frac{a_c^2\, \Psi_{aaa}(1,a_c,\Lambda_c)
}{2\Lambda_c\, \Psi_{a\Lambda}(1,a_c,\Lambda_c)}\, u^2\, ,
\label{cr200}
\end{equation}
or
\begin{equation}
\Lambda-\Lambda_c=A_1\, (a-a_c)^2\,, \label{cr240}
\end{equation}
where
\begin{equation}
A_1=-\frac{ \Psi_{aaa}(1,a_c,\Lambda_c) }{2
\Psi_{a\Lambda}(1,a_c,\Lambda_c)}\,. \label{cr250}
\end{equation}
One can see from Eq. (\ref{cr200}) that, close to the critical
point, $\delta ={\cal O} (u^2)$. That is why we could neglect the
term proportional to $\Psi_{\Lambda\Lambda}\, \delta^2 = {\cal
O}(u^4)$ in Eq.~(\ref{cr52}).

The coefficients $\Psi_{aaa}(1,a_c,\Lambda_c)$ and
$\Psi_{a\Lambda}(1,a_c,\Lambda_c)$ can be computed numerically,
see Appendix:
\begin{eqnarray}
 \Psi_{aaa}(1,a_c,\Lambda_c)&=&0.02434\dots\, ,
\label{cr220}\\
 \Psi_{a\Lambda}(1,a_c,\Lambda_c)&=&-0.001926\dots\, ,
\label{cr230}
\end{eqnarray}
so $A_1= 6.317\dots$.

Eq.~(\ref{cr240}) can be rewritten in terms of $\Lambda$ and $f$.
Using Eq. (\ref{cr21}), we expand $Z(x,a,\Lambda)$ near the
critical stripe solution $Z_c(x) \equiv Z(x,a_c,\Lambda_c)$ up to
the first order in $a-a_c$ (which suffices close to the critical
point) and obtain
\begin{equation}
f-f_c = f_a(a_c,\Lambda_c) \, (a-a_c)\,, \label{fvsa}
\end{equation}
where
\begin{equation}
f_a(a_c,\Lambda_c) = -\int\limits_0^1
\frac{\Psi_a(x,a_c,\Lambda_c)}{\bigl[Z_c(x)\bigr]^2\, F(Z_c(x))}\,
dx \,.\label{f_a}
\end{equation}
Evaluating this integral numerically, we obtain
$f_a(a_c,\Lambda_c) = -0.004396\dots$. As the result, Eq.
~(\ref{cr240}) can be rewritten as
\begin{equation}
\Lambda-\Lambda_c=A_2\, (f-f_c)^2\, ,
\label{cr260}
\end{equation}
where
\begin{equation}
A_2=A_1\, \bigl[f_a(a_c,\Lambda_c)\bigr]^{-2}= 3.268\dots\cdot
10^5\,.\label{cr270}
\end{equation}

Now we compute the spinodal line in the $f,P$-plane. Expanding Eq.
(\ref{Pr20a}) in the vicinity of $z= Z(1,a_c,\Lambda_c)$, we
obtain
\begin{equation}
P=
P_c+\Pi_{\psi}^{(c)}\Psi_\Lambda(1,a_c,\Lambda_c)(\Lambda-\Lambda_c)+\dots\,,
\label{cr280}
\end{equation}
where the higher-order terms are negligible, and
\begin{equation}
\Pi_{\psi}^{(c)}=
\left.\left(\frac{1}{F(z)}\frac{d\Pi}{dz}\right)\right|_{z=z[\Psi(1,a_c,\Lambda_c)]}
=  -2.969\dots\cdot 10^{-4}\,. \label{cr290}
\end{equation}
The negative value of $\Pi_{\psi}^{(c)}$ is a consequence of our
definitions of $z$ and $\psi$ and of the condition $(\partial
p/\partial n)_T>0$. Combining Eqs.~(\ref{cr280}) and (\ref{cr260})
we obtain:
\begin{equation}
P-P_c=-A_5(f-f_c)^2\,, \label{cr300}
\end{equation}
where $A_5=-A_2 \Psi_\Lambda(1,a_c,\Lambda_c) \Pi_{\psi}^{(c)}$. A
numerical calculation (see Appendix) gives
$\Psi_\Lambda(1,a_c,\Lambda_c)= 0.2203\dots$, therefore $A_5=
21.38\dots$. The spinodal asymptotics~(\ref{cr300})
and~(\ref{cr260}) are shown, together with the full spinodal line,
in Figs.~\ref{diagram2}a and~\ref{diagram2}b, respectively.

\subsection{Two-phase coexistence and binodal line in the vicinity of the critical point}

\subsubsection{Laterally non-uniform states}
For a steady state with a broken lateral symmetry the function
$\psi(x,y)$ depends on its two arguments. In a laterally infinite
system the asymptotics of $\psi(x,y)$ at $y \to \pm \infty$
correspond to two different stripe states. Therefore, it is
natural to replace the no-flux or periodic boundary conditions in
the lateral direction by the condition
\begin{equation}
\psi(x,y \to \pm \infty) = \psi_{\pm} (x)\,, \label{ii40}
\end{equation}
where $\psi_-(x) \neq \psi_+ (x)$. One way to solve the
problem~(\ref{i10})-(\ref{ii30}) and~(\ref{ii40}) is to introduce
an unknown function $a(y)$ so that
$$
\psi(0,y)=a(y)\, ,
$$
\begin{equation}
a(y\to\pm\infty)= a_{\pm}=\mbox{const} \label{bin174_0}
\end{equation}
with $a_{-}\ne a_{+}$. What equation should $a(y)$ satisfy close
to the critical point? Here we can look for $\psi(x,y)$ in the
form of a weakly and slowly modulated stripe state:
\begin{equation}
\psi(x,y)=\Psi[x,a(y),\Lambda]+\phi(x,y)\, , \label{bin20}
\end{equation}
where $a(y)=a_c[1+u(y)]$ is a slow function of $y$,  $\phi(x,y)$
is a small correction to $\Psi$, and a $y$-dependent order
parameter $u(y) \ll 1$ is introduced. We will see shortly that the
characteristic length scale of $a(y)$ (the domain wall width) is
of the order of $\delta^{-1/2} \sim u^{-1}$, while $\phi \sim
u^3$. Hence, every $y$-derivative introduces smallness of order
$u$. Making the Ansatz (\ref{bin20}) in Eq.~(\ref{i10}) and
neglecting terms of a higher order than $u^3$, we arrive at the
following linear problem for $\phi(x,y)$:
\begin{equation}
\partial_x^2\, \phi-\Lambda_c Q_\psi[\Psi_c(x)]\, \phi =
-a_c \, \Psi_a(x,a_c,\Lambda_c)\,\frac{d^2 u}{dy^2}\, ,
\label{bin60}
\end{equation}
\begin{equation}
\phi(0,y)=\partial_x\phi(0,y)=0\, . \label{bin70}
\end{equation}
The first boundary condition in Eq. (\ref{bin70}) elmiminates the
arbitrariness in the choice of $\phi(x,y)$, while the second one
follows from Eqs. (\ref{ii20}) and (\ref{cr20}). Additional
boundary conditions include
\begin{equation}
\label{wall_again} \Psi(1,a,\Lambda)+\phi(1,y) = const
\end{equation}
at the thermal wall [see Eq. (\ref{ii30})], and
\begin{equation}
\label{ii40a} \phi(x,y \to \pm \infty) = 0
\end{equation}
[see Eq. (\ref{ii40})]. Notice, however, that Eqs.
(\ref{wall_again}) and (\ref{ii40a}) do not enter the problem
(\ref{bin60}) and (\ref{bin70}) for $\phi (x,y)$.
Equation~(\ref{bin60}) can be solved by separation of variables:
\begin{equation}
\phi(x,y)=a_c\, \Phi(x)\,\frac{d^2 u(y)}{dy^2}\, . \label{bin80}
\end{equation}
The function $\Phi(x)$ is the solution of the following problem:
\begin{equation}
\Phi^{\prime\prime}-\Lambda_c Q_\psi[\Psi_c(x)]\, \Phi =
-\Psi_a(x,a_c,\Lambda_c)\, , \label{bin90}
\end{equation}
\begin{equation}
\Phi(0)=\Phi^{\prime}(0)=0\, ,\label{bin100}
\end{equation}
which again belongs to the class of equations (\ref{gen_form}).
The solution (see Appendix) is
$$\Phi(x)=
\frac{1}{\Lambda_c Q(a_c)}\Biggl[ \Psi_a
(x,a_c,\Lambda_c)\int\limits_0^x \Psi^{\prime} \Psi_a \, d\xi$$
\begin{equation}
-\Psi^{\prime}(x,a_c,\Lambda_c)\int\limits_0^x \Psi_a^2 \,
d\xi\Biggr]\, . \label{bin110}
\end{equation}
where the functions $\Psi^{\prime}$ and $\Psi_a$ under the
integrals over $\xi$ have arguments $\xi,a_c$ and $\Lambda_c$. Now
we impose the boundary condition (\ref{wall_again}):
\begin{equation}
\label{wall_again2} \Psi(1, a, \Lambda)+a_c \Phi(1) \frac{d^2u}{d
y^2} = const\,,
\end{equation}
As the second term in the right side of Eq. (\ref{wall_again}) is
of order $u^3$, we should keep terms up to $u^3$ in the expansion
of $\Psi(x,a,\Lambda)$ near the critical point $a_c,\Lambda_c$.
This expansion has the same form as Eq. (\ref{cr52}), with the
only difference that now $u$ depends on $y$. Evaluating this
expansion at $x=1$ and using the definitions of the critical point
[Eqs. (\ref{cr40}) and (\ref{cr50})], we arrive at the desired
equation for $u(y)$:
\begin{equation}
u\, \delta- A_3\, u^3 + A_4\, \frac{d^2 u}{dy^2}=\alpha = const\,
, \label{bin120}
\end{equation}
where
\begin{equation}
A_3=-\frac{a_c^2\Psi_{aaa}(1,a_c,\Lambda_c)}{6\Lambda_c
\Psi_{a\Lambda}(1,a_c,\Lambda_c)}= \frac{a_c^2 A_1}{3 \Lambda_c}\,
= 0.26485\dots\, , \label{bin130}
\end{equation}
\begin{equation}
A_4=\frac{\Phi(1)}{\Lambda_c \Psi_{a\Lambda}(1,a_c,\Lambda_c)}\, =
0.5212\dots\, , \label{bin140}
\end{equation}
and [see Eq.~(\ref{bin110})]
\begin{equation}
\Phi(1)=-\frac{\Psi^{\prime}(1,a_c,\Lambda_c)}{\Lambda_c Q(a_c)}
\int\limits_0^1  \Psi_a^2\, dx = -0.3457\dots\, . \label{bin150}
\end{equation}

\subsubsection{Domain wall and the binodal line}

One integration of Eq.~(\ref{bin120}) yields:
\begin{equation}
A_4\, \left(\frac{du}{dy}\right)^2= \frac{A_3}{2}\,
\left(u^2-\frac{\delta}{A_3}\right)^2 +2\alpha u+\beta\, ,
\label{bin190}
\end{equation}
where $\beta =const$. To find the constants $\alpha$ and $\beta$,
we are using the boundary conditions in the lateral direction. In
an infinite system these are
\begin{equation}
u(y\to\pm\infty)= u_{\pm}=\mbox{const} \label{bin174}
\end{equation}
with $u_{-}\ne u_{+}$. Equations (\ref{bin190}) and (\ref{bin174})
appear in numerous problems of domain wall structure, see for
example Ref. \cite{LL_electr}. The boundary conditions
(\ref{bin174}) yield $\alpha=\beta=0$, so the domain wall solution
is
\begin{equation}
u=\pm\sqrt{\frac{\delta}{A_3}}\, \tanh \left[\sqrt{\frac{\delta}{2
A_4}}\, \left(y-y_0\right)\right]\, , \label{bin210}
\end{equation}
where $\pm$ refers to two possible orientations, and the arbitrary
constant $y_0$ describes the position of the domain wall. The
domain wall solution exists only if $\delta>0$, that is
$\Lambda>\Lambda_c$. Eq.~(\ref{bin210}) confirms our assumption
that the characteristic width of the domain wall is ${\cal
O}(\delta^{-1/2})$. Returning for a moment to the physical
(unscaled) variables, we see that the domain wall width is ${\cal
O} (L_x/\delta^{1/2})$ which is much larger than $L_x$.

The values of the order parameter far from the domain wall are
\begin{equation}
u_\pm=\frac{a_{\pm}}{a_c}-1=\pm\sqrt{\frac{\delta}{A_3}}\, .
\label{bin200}
\end{equation}
Equation~(\ref{bin200}), rewritten in terms of $a$ and $\Lambda$,
defines the binodal (coexistence) line:
\begin{equation}
\Lambda-\Lambda_c=\frac{A_1}{3}\, (a-a_c)^2\,, \label{bin220}
\end{equation}
or, in terms of $f$ and $\Lambda$,
\begin{equation}
\Lambda-\Lambda_c=\frac{A_2}{3}\, (f-f_c)^2\, .
\label{bin230}
\end{equation}
Compare these expressions with Eqs.~(\ref{cr240})
and~(\ref{cr260}) for the spinodal.

The binodal line can be also expressed in terms of $f$ and $P$.
The derivation almost coincides with that for the spinodal line,
see the end of Sec.~\ref{spcr}. The only difference is that now we
substitute into Eq.~(\ref{cr280}) the binodal
relation~(\ref{bin230}) rather than the spinodal relation
(\ref{cr260}). The result is
\begin{equation}
P-P_c=-\frac{A_5}{3}\, (f-f_c)^2\, . \label{bin240}
\end{equation}

The physical meaning of the binodal asymptotics~(\ref{bin240})
and~(\ref{bin230}) is straightforward. Firstly, the two coexisting
stripes with $f=f_1$ and $f=f_2$ have equal pressures:
$P(f_1)=P(f_2)$. Secondly, close to the critical point, $f_1$ and
$f_2$ are symmetric with respect to $f_c$, that is $(f_1+f_2)/2 =
f_c$. The binodal asymptotics~(\ref{bin240}) and~(\ref{bin230})
are depicted in Figs.~\ref{diagram2}a and ~\ref{diagram2}b,
respectively.

Note that our results for the spinodal and binodal lines [see Eqs.
(\ref{cr260}) and (\ref{bin230})] are consistent with the results
obtained by Soto \textit{et al.} \cite{Argentina1,Argentina2} in
the framework of their phenomenological ``van der Waals equation".
Indeed, our Eqs. (\ref{cr260}) and (\ref{bin230}) have the same
quadratic forms as those obtainable from the van der Waals
equation \cite{Argentina1, Argentina2}. Furthermore, the ratio of
the coefficients of the spinodal and binodal lines (\ref{cr260})
and (\ref{bin230}) equals three, in agreement with  what the van
der Waals equation predicts. (Note that the coefficients
themselves of the van der Waals equations have not been derived
yet.) In addition, Soto \textit{et al.} \cite{Argentina2} reported
a binodal line found in molecular dynamics simulations of this
system for a moderate value of the parameter $\varepsilon = d/L_x
= 0.01$. See Sec. 5 for a discussion of the role of this
parameter.

Soto \textit{et al.} \cite{Argentina2} also suggested an
interpretation of the binodal line in terms of the ``Maxwell's
construction". By analogy with the classical van der Waals gas
\cite{Huang}, Maxwell's construction can be written as
\begin{equation}
\int\limits_{f_1}^{f_2} \frac{P(f,\Lambda)-P(f_1,\Lambda)}{f^2}
\,d f =0\,, \label{maxwell1}
\end{equation}
where the factor $f^2$ in the denominator comes from equality $d V
= d\, (1/f) = - df/f^2$, where $V=1/f$ is the (scaled) specific
volume. How does Eq. (\ref{maxwell1})  compare to our result
(\ref{bin240})? Consider expression (\ref{Pr20a}) for
$P(a,\Lambda)$. Close to the critical point, it can be expanded in
the powers of $u$ and $\delta$:
\begin{equation}
P=P_c+\Lambda_c P_\Lambda \delta + a_c\Lambda_c P_{a\Lambda} u
\delta+\frac{1}{6} a_c^3 P_{aaa} u^3+\dots\,,  \label{mx10}
\end{equation}
where all the derivatives of $P$ are evaluated at the critical
point. Going over from $u$ to $f-f_c$, one obtains
\begin{eqnarray}
P=P_c+\Lambda_c P_\Lambda \delta+ \frac{\Lambda_c P_{a\Lambda}
\delta}{f_a}(f-f_c) \nonumber \\
+ \,\frac{P_{aaa}}{6f_a^3}(f-f_c)^3+\dots\,, \label{mx20}
\end{eqnarray}
where the numerical coefficient $f_a$ is given by Eq. (\ref{f_a}),
and the dots stand for terms of the order of $(f-f_c)^4$ and
higher. At $\delta>0$ (when phase separation occurs), the
coefficient in front of $f-f_c$ is negative, while that in front
of $(f-f_c)^3$ is positive.  Note also that, in the vicinity of
the critical point, $\delta = {\cal O} (f-f_c)^2$. Now we
substitute Eq. (\ref{mx20}), with the higher-order terms
neglected, into Eq. (\ref{maxwell1}). In this order of
perturbation theory one should put $f=f_c$ in the denominator of
the integrand. As the result, Eq. (\ref{maxwell1}) reduces to the
simple relation $(f_1+f_2)/2 = f_c$ obtained above. Of course, the
reason for this agreement is the closeness to the critical point.
In this sense, the Maxwell's construction (\ref{maxwell1}) does
not give anything new. Moreover, many \textit{different}
constructions, for example,
\begin{equation}
\int\limits_{f_1}^{f_2} \left[P(f,\Lambda)-P(f_1,\Lambda)\right]
\,df =0\,, \label{maxwell}
\end{equation}
are equally applicable in the vicinity of the critical point. Of
course, the Maxwell's construction would be valuable if it were
shown to be true \textit{far} from the critical point. At present
there is no reason to believe this is the case, and the
theoretical form of the binodal line far from the critical point
remains unknown.

\section{Finite size effects: marginal stability and
bifurcations}

\subsection{Marginal stability surface}

\begin{figure}
\includegraphics[width=7.0cm,clip=]{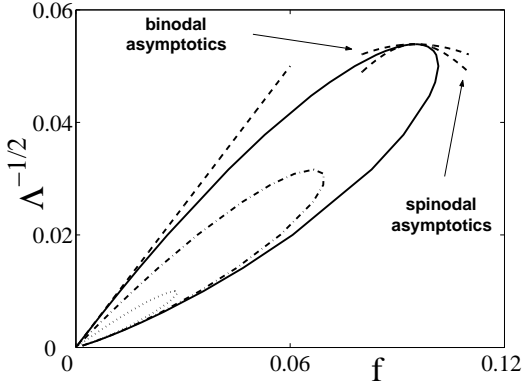}
\caption{Cross-sections of the marginal stability surface by three
planes: $\Delta=\infty$ (the solid line), $\Delta=1.788$ (the
dash-dotted line) and $\Delta=0.523$ (the dotted line). The dashed
lines show the dilute-limit asymptote $f\Lambda^{1/2}=1.199 7...$,
obtained analytically \cite{KM}, and the spinodal and binodal
asymptotics near the critical point.} \label{etta_f}
\end{figure}

The results presented in Sections 2 and 3 are valid in the limit
of $\Delta \to \infty$. Already in the first paper \cite{LMS} on
the phase separation instability  it was found that the
instability is suppressed when the aspect ratio of the box
$\Delta=L_y/L_x$ is less than a threshold value
$\Delta_*(f,\Lambda)$. The physical mechanism of suppression is
heat conduction in the lateral direction which tends to erase the
lateral temperature (and, therefore, density) inhomogeneity. How
to generalize the spinodal line, obtained for $\Delta \to \infty$,
to finite $\Delta$-s?  Let us consider a small sinusoidal density
perturbation, in the lateral direction, around a stripe state. The
fastest growing (or the slowest decaying) perturbation is the one
with the longest wavelength, compatible with the boundary
conditions in the lateral direction \cite{LMS,Brey2,KM}. Consider
a three-dimensional parameter space $(f,\Lambda,\Delta)$ and
define in it a two-dimensional \textit{marginal stability} surface
${\cal F} (f,\Lambda,\Delta)=0$. By definition, at any point on
this surface the growth rate of the longest perturbation is equal
to zero. The marginal stability surface represents a natural
generalization of the spinodal line. Importantly, this definition
reduces to that of the spinodal line as $\Delta \to \infty$ [see
Ref. \cite{KM} and Eq. (\ref{mstab}) below]. The marginal
stability surface can be computed by linearizing Eq. (\ref{i10})
around the stripe state and solving the resulting linear
eigenvalue problem. Calculations of this kind were done previously
for large $\Lambda$ far from the critical point
\cite{LMS,Brey2,KM}. Three typical cross-sections of the marginal
stability surface are shown in Fig.~\ref{etta_f}. As expected, the
instability region shrinks as $\Delta$ goes down. As the result,
the critical point moves toward smaller $f$-s and larger
$\Lambda$-s as $\Delta$ decreases. The monotonic dependence of the
critical point position on $\Delta$ is explained by the monotonic
increase of the lateral heat conduction as $\Delta$ decreases.

\begin{figure}
\includegraphics[width=7.0cm,clip=]{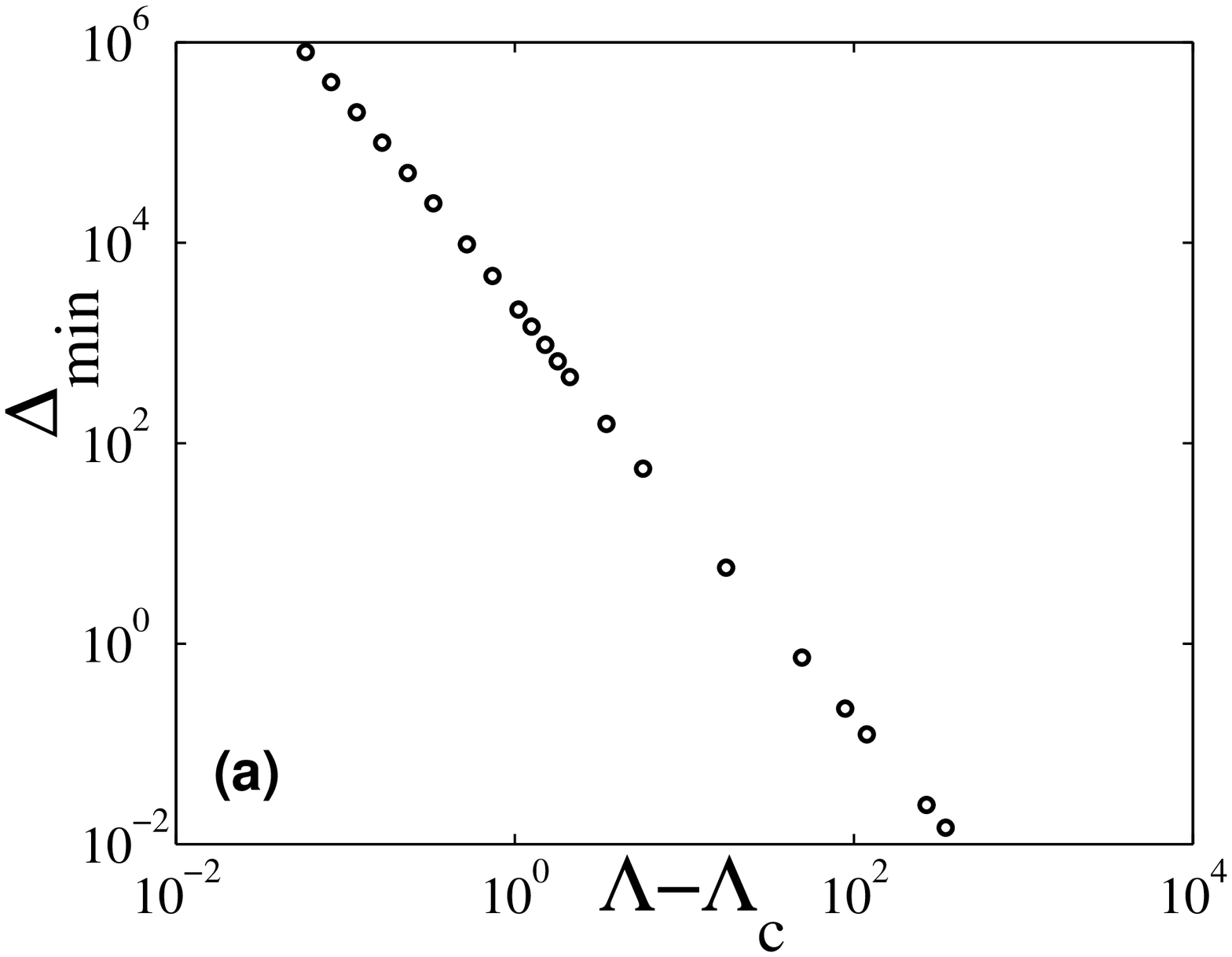}
\includegraphics[width=7.0cm,clip=]{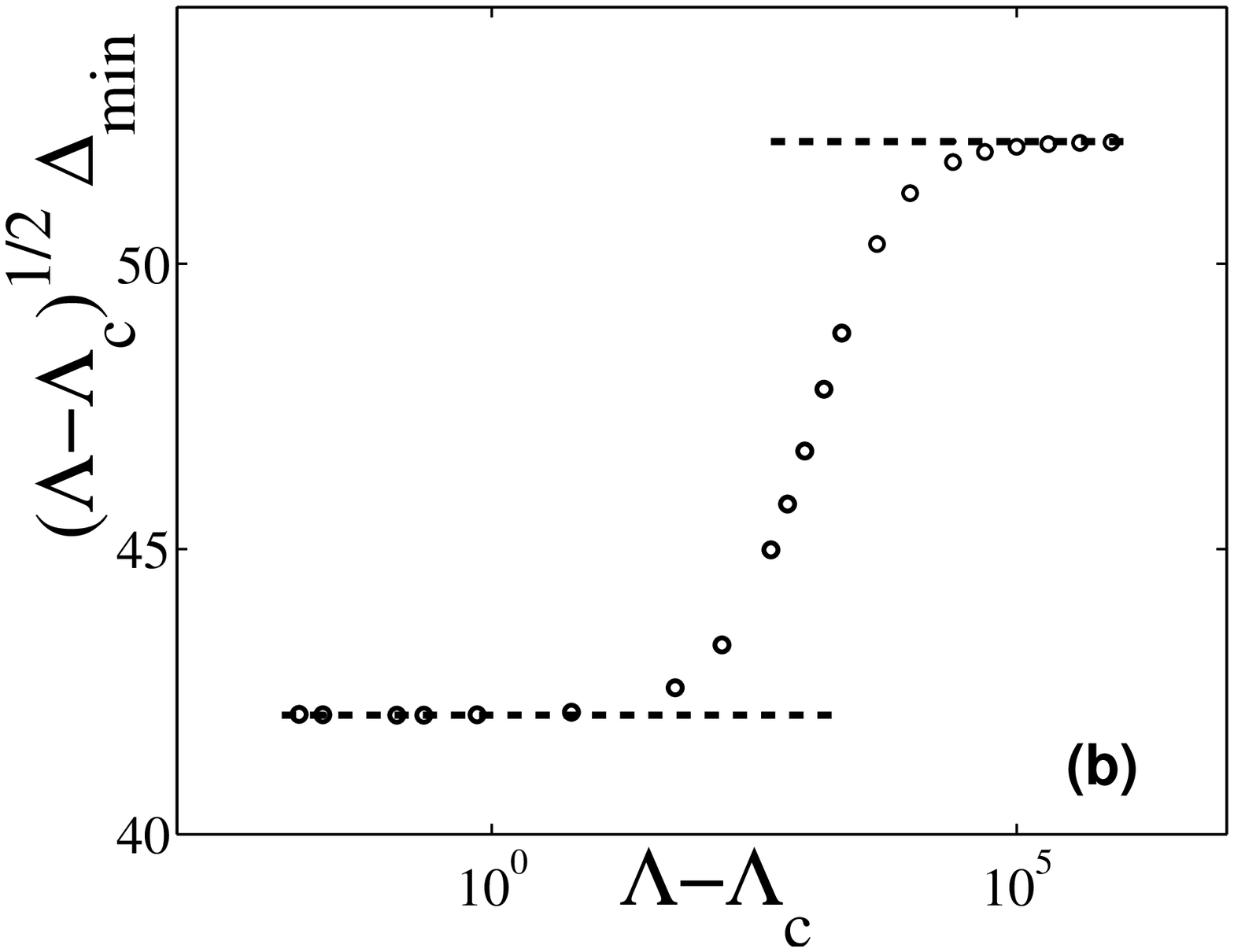}
\caption{The dependence of the threshold value of the aspect ratio
$\Delta$ for the suppression of the phase separation on the
parameter $\Lambda-\Lambda_c$ (a). The phase separation
instability occurs when $\Delta > \Delta_{min} (\Lambda)$. Figure
b shows that the apparent ``straight line" in figure a actually
includes two different asymptotes and a crossover between them.
The two asymptotes are shown by the dashed lines: the $\Lambda \gg
1$ asymptote, $\Delta_{min}(\Lambda) = 52.14...\,\Lambda^{-1/2}$,
and the asymptote near the critical point, $\Delta_{min} =
\pi\,(\Lambda_c\,A_4)^{1/2}\,(\Lambda-\Lambda_c)^{-1/2}$, where
$A_4=0.5213 \dots$. The results of numerical marginal stability
analysis are denoted by circles.} \label{deltamin}
\end{figure}

The threshold value of the aspect ratio $\Delta_*(\Lambda, f)$ has
a minimum at some $f$ \cite{LMS,KM}. The respective minimum value
$\Delta_{min}$ depends only on $\Lambda$.  In this work we
performed a systematic investigation of this dependence. We found
that $\Delta_{min}$ goes down monotonically as the parameter
$\Lambda-\Lambda_c$ is positive and increases. Though this
monotonic decrease looks like a power law in the log-log plot, see
Fig.~\ref{deltamin}a, it is actually not. Figure~\ref{deltamin}b
shows the same dependence on a different scale. Two different
asymptotes are clearly seen. The first of them, at $\Lambda \gg
1$, was obtained previously: $\Delta_{min}(\Lambda) = A
\,\Lambda^{-1/2}$, where $A = 52.14\dots$ \cite{LMS,KM}. In this
regime the eigenfunction of the marginal stability problem is
exponentially localized at the elastic wall $x=0$. The second
asymptote is valid close to the critical point
$\Lambda=\Lambda_c$, where $\Delta_{min}$ diverges: $\Delta_{min}
= 42.085\dots\, (\Lambda-\Lambda_c)^{-1/2}$. This asymptote is
derived analytically in the next subsection.

An additional interesting issue is the change of bifurcation
character, predicted by a weakly nonlinear analysis of the steady
state problem close to the marginal stability surface. At fixed
$\Lambda$, the bifurcation is supercritical on an interval
$f_{-}(\Lambda)<f<f_{+}(\Lambda)$ which lies within the spinodal
interval $(f_1,f_2)$. On the intervals $f_1<f<f_{-}$ and
$f_{+}<f<f_2$ the bifurcation is subcritical \cite{LMS2,MPSS}. The
next subsection addresses the finite size effects in the vicinity
of the critical point, where everything can be calculated
analytically.

\subsection{Steady states and bifurcation types close to the
critical point}

This subsection addresses two-dimensional steady states in a
laterally finite system close to the critical point. As
$\Delta_{min}$ diverges at the critical point [see
Eq.~(\ref{threshold_min}) below], we assume that $\Delta$, though
finite, is very large.  The starting point here is the same
Eq.~(\ref{bin120}), but on a finite interval $-\Delta/2\leq y\leq
\Delta/2$, so we replace the boundary conditions ~(\ref{bin174})
by the no-flux conditions
\begin{equation}
\left.\frac{du}{dy}\right|_{y=-\frac{\Delta}{2}}=\left.\frac{du}{dy}\right|_{y=\frac{\Delta}{2}}=0
\,. \label{fin20}
\end{equation}
Periodic boundary conditions can be treated in a similar manner.
Integrating Eq.~(\ref{bin120}) over $y$ from $-\Delta/2$ to
$\Delta/2$ and using the boundary conditions (\ref{fin20}), we
determine $\alpha$ and rewrite Eq.~(\ref{bin120}) as
\begin{equation}
A_4\, \frac{d^2 u}{dy^2}+ u\, \delta- A_3\, u^3 -\left\langle u\,
\delta - A_3\, u^3\right\rangle=0\, , \label{fin22}
\end{equation}
where $\langle\dots\rangle$ denotes spatial averaging:
$$
\langle
\dots\rangle=\frac{1}{\Delta}\int\limits_{-\Delta/2}^{\Delta/2}
(\dots)\, dy\, .
$$
Introduce the rescaled coordinate $y_*=y/\Delta$, order parameter
$u_*=u \,\Delta$ and control parameter $\delta_*=\Delta^2 \,
\delta$. Equation (\ref{fin22}) keeps its form in the rescaled
variables,
\begin{equation}
A_4\, \frac{d^2 u_*}{dy^2}+ u_*\, \delta_*- A_3\, u_*^3 -\langle
u_*\rangle\, \delta_* + A_3\, \langle u_*^3\rangle=0\, ,
\label{fin24}
\end{equation}
while the boundary conditions become
\begin{equation}
\frac{d u_*}{dy}(y_*=-1/2)=\frac{d u_*}{dy}(y_*=1/2)=0\, .
\label{fin26}
\end{equation}
As the aspect ratio $\Delta$ drops out in these variables, a
universal description can be obtained.

\begin{figure}[ht]
\includegraphics[width=7.0cm,clip=]{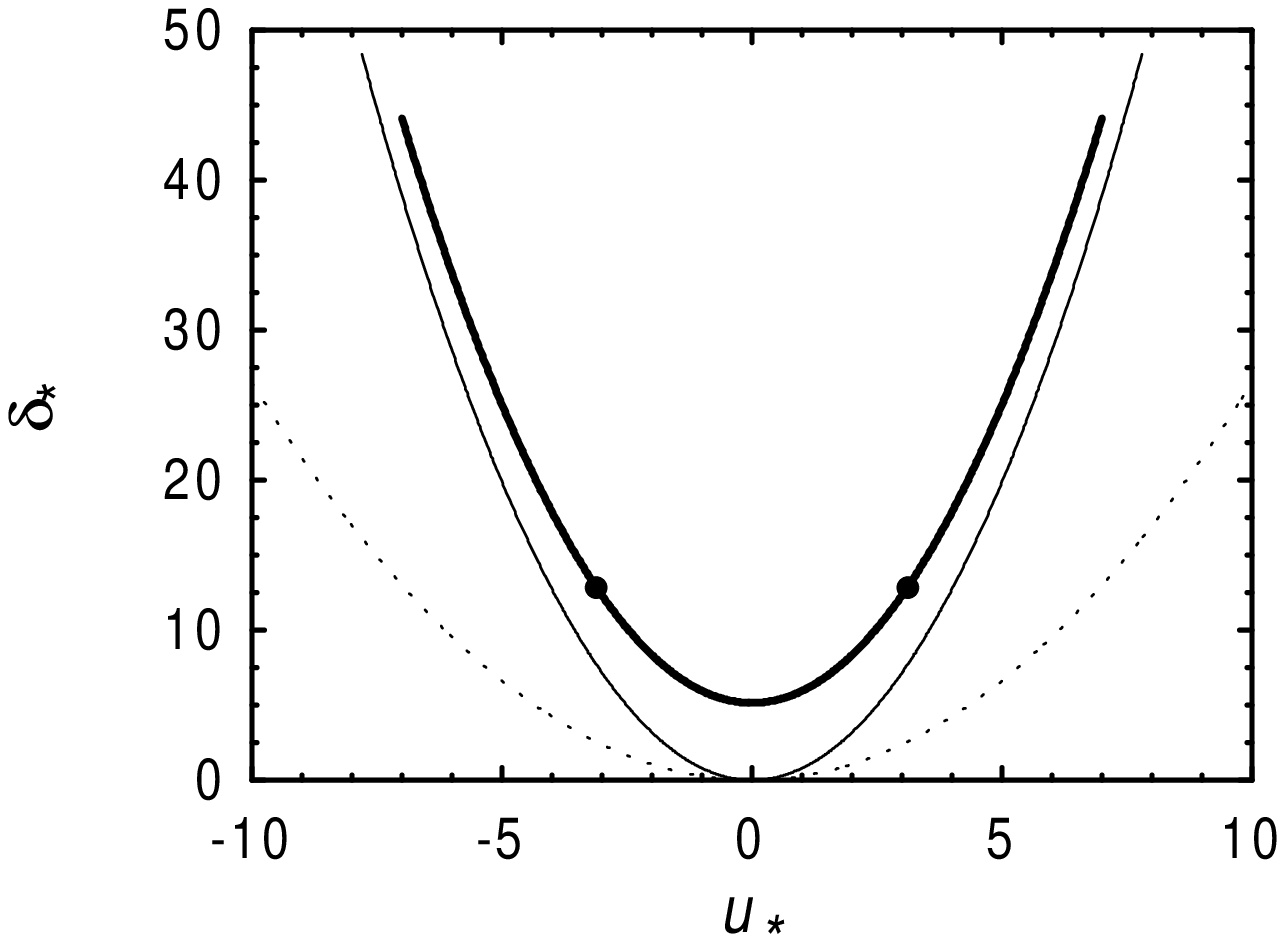}
\caption{The marginal stability curve for a laterally finite
system (the thick solid line) is shown together with the spinodal
and binodal lines for an infinite system (the thin solid line and
the thin dotted line, respectively). This is made possible by
using the rescaled order parameter $u_*$ and control parameter
$\delta_*$. The two black circles show the points
(\protect\ref{co100}) and ~(\protect\ref{co110}), where the
bifurcation changes its character from the supercritical (between
the circles) to the subcritical (outside).} \label{figS2}
\end{figure}

Obviously, any $y$-independent state $u_*=const$ solves the
problem~(\ref{fin24})-(\ref{fin26}) (a $y$-independent state is
nothing but a stripe state). What is the condition for the
appearance of (weakly) $y$-dependent solutions? When a
$y$-dependent solution does appear, what is the type of
bifurcation? To address these questions, we seek for a weakly
$y$-dependent solution in the form
\begin{equation}
u_*=\langle u_*\rangle+a_1\sin\pi y_*+a_2\cos 2\pi y_*+\dots \, ,
\label{ma2}
\end{equation}
We substitute Eq. (\ref{ma2}) into Eq. (\ref{fin24}) and treat the
terms originating from $a_2 \cos 2 \pi y_*$ as small corrections.
Expanding up to ${\cal O} (a_1^3)$, we obtain the following two
algebraic equations:
$$
\delta_* - \pi^2 A_4 - 3 A_3 \,\langle u_* \rangle^2 -
\frac{3}{4}\, A_3 a_1^2 + 3 A_3 \langle u_* \rangle \, a_2 = 0\,,
$$
\begin{equation}\label{additional}
\frac{3}{2}\, A_3 \langle u_* \rangle \,a_1^2+ \left(\delta_*-4
\pi^2 A_4 -3 A_3 \langle u_* \rangle \right) a_2 = 0 \,.
\end{equation}
Putting here $a_1=a_2=0$, we obtain the marginal stability
condition
\begin{equation}
\delta_*=3A_3\, u_*^2+\pi^2\, A_4\,, \label{ma60}
\end{equation}
where we have omitted the spatial averaging of $u_*$, as it
becomes trivial on the marginal stability curve. The marginal
stability curve is shown, as the thick solid line, in
Fig.~\ref{figS2}. Now we consider nonzero amplitudes $a_1$ and
$a_2$ in Eq. (\ref{additional}) and eliminate $a_2$ in favor of
$a_1$. Above the marginal stability curve (\ref{ma60}), but close
to it, we obtain the following equation for the bifurcation curve:
\begin{equation}
\frac{\delta_*-3A_3\,  \langle u_*\rangle^2-\pi^2A_4}{3A_3}=
\frac{a_1^2}{4} \left(1-\frac{2A_3  \langle
u_*\rangle^2}{\pi^2A_4}\right) . \label{ma64}
\end{equation}
One can see that, on the marginal stability curve, the bifurcation
is either supercritical (when the term in the parentheses in the
right hand side of this equation is positive), or subcritical
(when this term is negative). The change of character of the
bifurcation occurs at the points
\begin{eqnarray}
\langle u_* \rangle &=&\pm\left(\frac{\pi^2A_4}{2A_3}\right)^{1/2}
=
 \pm \, 3.116\dots\,,
\label{co100}\\
\delta_*&=&\frac{5}{2}\, \pi^2\, A_4 \simeq 12.86\dots \, .
\label{co110}
\end{eqnarray}
which lie on the marginal stability curve~(\ref{ma60}), see
Fig.~\ref{figS2}.

Note that the spinodal and binodal lines for an infinite system
can be expressed through $u_*$ and $\delta_*$: $\delta_*=3A_3\,
u_*^2$ and $\delta_*=A_3\, u_*^2$, respectively. This makes it
possible to show all the curves (the spinodal and binodal for an
infinite system  and the marginal stability curve for a finite
system) on the same plot, see Fig.~\ref{figS2}.

Going back from the rescaled variables $\delta_*$ and $u_*$ to
$\Lambda$ and $a$, we obtain for the marginal stability curve:
\begin{equation}
\label{mstab} \Lambda - \Lambda_c = A_1 \,(a-a_c)^2 +\frac{\pi^2
A_4 \Lambda_c}{\Delta^2}\,,
\end{equation}
or, in the variables $f,\Lambda$:
\begin{equation}
\label{mstab1} \Lambda - \Lambda_c = A_2 \,(f-f_c)^2 +\frac{\pi^2
A_4 \Lambda_c}{\Delta^2}\,.
\end{equation}
The physical meaning of this result becomes transparent when one
compares it with Eq. (\ref{cr260}) for the spinodal in the
infinite system.

Solving Eq. (\ref{mstab1}) for $\Delta$, we can find the threshold
value of the aspect ratio $\Delta_*$ at given $f$ and $\Lambda$:
\begin{equation}
\label{threshold} \Delta_* = \frac{\pi\,A_4^{1/2}
\Lambda_c^{1/2}}{\left[\Lambda-\Lambda_c -A_2
(f-f_c)^2\right]^{1/2}}\,.
\end{equation}
The phase separation instability occurs at $\Delta>\Delta_*$. One
can see that $\Delta_*$ diverges on the spinodal line. The minimum
value of $\Delta_*$ at a given $\Lambda$ corresponds to $f=f_c$:
\begin{equation}
\label{threshold_min} \Delta_{min} = \frac{\pi\,A_4^{1/2}
\Lambda_c^{1/2}}{(\Lambda-\Lambda_c )^{1/2}}\,.
\end{equation}
This quantity diverges at the critical point, as the lower
asymptote in Fig.~\ref{deltamin} shows.

Finally, the two points of change of the bifurcation character,
[Eqs. (\ref{co100}) and (\ref{co110})], form a line,
\begin{equation}
\label{change1} \Lambda-\Lambda_c= \frac{5}{3} \,A_1\, (a-a_c)^2
\end{equation}
in the plane $a,\Lambda$, which corresponds to the line
\begin{equation}
\label{change2} \Lambda-\Lambda_c= \frac{5}{3} \,A_2\,
(f-f_c)^2\,,
\end{equation}
in the plane $f,\Lambda$. This line lies within the spinodal
interval of the problem corresponding to $\Delta \to \infty$.

\section{Summary and Discussion}

In this work we have employed granular hydrostatics to determine
the phase diagram of a prototypical driven granular gas which
exhibits spontaneous symmetry breaking and van der Waals-like
phase separation. We determined the spinodal line and the critical
point of the phase separation. We computed the spinodal and
binodal (coexistence) lines close to the critical point. Effects
of finite lateral size of the confining box have been also
addressed. These include determining the line of change of the
bifurcation character from supercritical to subcritical.

The shape of the binodal (coexistence) line \textit{far} from the
critical point is still unknown. To handle this problem within the
framework of the hydrostatic theory one needs to solve the
nonlinear Poisson equation (\ref{i10}) in a laterally infinite
box. Most likely, this can only be done numerically, in a
sufficiently long box. It is clear, however, that the binodal line
in the variables $(a, \Lambda)$ is solely determined by the
function $Q(\psi)$ which encapsulates all the necessary
information about the equation of state, heat conductivity and
inelastic energy loss.

It is worth mentioning that our \textit{qualitative} results (the
phase separation instability, the existence of the critical point
and spinodal and binodal lines, the suppression of the instability
by the lateral heat conduction and the change of bifurcation
character) are robust and independent of the small details of the
constitutive relations. Furthermore, most of our analytical
results are presented in quite a general form which only assumes a
Navier-Stokes structure of the hydrodynamic equations (in the
limit of nearly elastic collisions). We used the Enskog-type
constitutive relations \cite{Jenkins} only for computing numerical
factors (and for the computations far from the critical point).
Obviously, the accuracy of our quantitative results cannot be
better then the accuracy of the Enskog-type constitutive
relations. Therefore, a 10-15 percent error margin should not be
surprising.

There is an important additional aspect that can become crucial
when comparing the hydrodynamic theory with molecular dynamic
simulations. As was already mentioned, for the hydrodynamics to be
quantitatively accurate, the mean free path of the particles must
be much less than any hydrodynamic length scale. 
It is well known that, for a system with a thermal wall, the
leading correction to hydrodynamics enters the \textit{boundary
condition} in the form of a temperature jump, proportional to the
ratio of the mean free path and  the characteristic hydrodynamic
length scale \cite{temp_jump,Stat2}. Indeed, within the Knudsen
layer, whose size is comparable to the mean free path,
hydrodynamics is inapplicable, while the particle velocity
distribution significantly deviates from Maxwell distribution: the
temperature of the incoming particles is less than that of the
outgoing particles [as the outgoing particles have their normal
velocity randomized according to a Maxwell distribution with a
(fixed) higher temperature]. The effective temperature at the
wall, for the purpose of a hydrodynamic description in the bulk
(that is, outside the Knudsen layer), is always less than $T_0$
\cite{temp_jump,Stat2}, as is indeed observed in MD simulation of
this system \cite{Grossman}. As the result, the pressure of the
system is reduced (notice that the density does not change in this
order; it can change only in the next, Burnett order). The
pressure reduction will obviously cause a shift of the critical
point. For our numerical results to be sufficiently accurate the
parameter $\varepsilon = d/L_x$ must be very small, so that the
mean free path is indeed much less than the system dimensions. As
$\Lambda \sim q\, \varepsilon^{-2} =const$ (for example, at the
critical point $\Lambda=\Lambda_c$), a very small $\varepsilon$
implies an extremely small inelasticity $q$. Despite the severe
limitation intrinsic to it, the nearly elastic case is
conceptually important, just because hydrodynamics is supposed to
give here a quantitatively accurate leading order theory.

An interesting direction for future work is the phase separation
\textit{dynamics}. Here the hydrostatic equations should give way
to the full set of hydrodynamic equations. Close to the critical
point, however, a reduced description of the dynamics should be
possible. It has been suggested that such a description is
provided by the ``van der Waals equation"
\cite{Argentina1,Argentina2}. Though the van der Waals equation
does capture qualitatively much of the phenomenology of the phase
separation as seen in the MD simulations \cite{Argentina1,
Argentina2}, its systematic derivation from the equations of
granular hydrodynamics is still lacking, and the coefficients of
the normal form are yet unknown. Importantly, our hydrostatic
results close to the critical point, encapsulated in Eq.
(\ref{bin120}), fully agree with those predicted from the van der
Waals normal equation. Furthermore, Eq. (\ref{bin120}) provides
quantitative relations between the yet unknown coefficients of the
van der Waals equation.

Another avenue of future work requires going beyond hydrodynamic
description, as it includes two types of fluctuations in this
system. The first of them was observed in molecular dynamic
simulations inside the spinodal region at $\Lambda = 11 \,050$
(that is, far from the critical point) in a wide region of aspect
ratios around the finite-size threshold value $\Delta_*$
\cite{MPSS}. It was found that these fluctuations dominate the
dynamics of the system, so they were called giant fluctuations.
The second type of fluctuations is expected to occur, by analogy
with the classical van der Waals phase transition, in a close
vicinity of the critical point. The fluctuations in these two
regimes should be describable in the framework of ``fluctuating
hydrodynamics" of Landau and Lifshitz \cite{Stat2}, generalized to
granular gases in the limit of nearly elastic collisions.
Fluctuating hydrodynamics is a Langevin-type theory which takes
into account the discrete character of particles by adding
delta-correlated noise terms in the momentum and energy equations
\cite{Stat2}. The fluctuations appear in this approach as a
hydrodynamic response of the system to the Langevin noise.
Unfortunately, this exciting direction of work is hindered by the
fact that the Langevin term, which accounts for the discreteness
of the inelastic energy loss in the energy equation, has yet to be
calculated \cite{MPSS}.

Finally, it was assumed throughout the paper that the granulate is
driven by a thermal wall. In experiment a rapidly vibrating wall
is usually used. Though qualitatively similar, the phase diagram
of the case of a rapidly vibrating wall can be quantitatively
different \cite{LMS,Brey2,LMS2}.

\begin{acknowledgments} This research was partially supported by
the Israel Science Foundation (Grant No. 180/02), by the Russian
Foundation for Basic Research (Grant No. 02-01-00734), and by the
Forchheimer Foundation.
\end{acknowledgments}


\section*{APPENDIX. COMPUTING THE COEFFICIENTS}

\begin{table*}[!]
\begin{center}
\begin{ruledtabular}
\begin{tabular}{lcccc}
$w(x)$                                &  $S(x)$        &            \multicolumn{2}{c}{Initial  conditions at $x=0$}  \\
\multicolumn{2}{c}{}& ~~~~~$w(0)$~~~~~   & ~~~$w^{\prime}(0)$~~~
\\ \colrule
$\Psi^\prime(x,a_c,\Lambda_c)\;$        &0                                                                                                                 & 0 & $\Lambda_c\, Q\left(a_c\right)$\\
$\Psi_a(x,a_c,\Lambda_c)\;$             &0                                                                                                                 & 1 & 0\\
$\Psi_{\Lambda}(x,a_c,\Lambda_c)\;$    &$Q \left(\Psi_c\right)$                                                                                 & 0 & 0\\
$\Psi_{aa}(x,a_c,\Lambda_c)\;$         &$\Lambda_c\, Q_{\psi\psi}\left(\Psi_c\right)\, \Psi_a^2$                                                          & 0 & 0\\
$\Psi_{\Lambda a}(x,a_c,\Lambda_c)\;$   &~~$\Lambda_c\, Q_{\psi\psi}\left(\Psi_c\right)\, \Psi_\Lambda\, \Psi_a +Q_{\psi}\left(\Psi_c\right)\, \Psi_a$~~~& 0 & 0\\
$\Psi_{aaa}(x,a_c,\Lambda_c)\;\;\;\;\;$      &$3 \Lambda_c\, Q_{\psi\psi}\left(\Psi_c\right) \Psi_a \Psi_{aa}+  \Lambda_c\, Q_{\psi\psi\psi}\left(\Psi_c\right)\, \Psi_a^3$     & 0 & 0\\
$\Phi(x,a_c,\Lambda_c)\;$              &$-\Psi_a$                                                                                                         & 0 & 0\\
\end{tabular}
\end{ruledtabular}
\end{center}
\caption{ The source terms $S(x)$ and the initial conditions at
$x=0$ for Eq.~(\protect\ref{fo10}), for each of the auxiliary
functions shown in the first column. The functions in the first
and second columns have the same arguments. \label{table}}
\end{table*}

Consider the stripe solution $ \Psi(x,a,\Lambda$). The expansions
in the vicinity of the critical point, used throughout the paper,
include several derivatives of this function which need to be
evaluated. These are $\Psi^\prime$, $\Psi_a$, $\Psi_{aa}$,
$\Psi_{\Lambda}$, $\Psi_{\Lambda a}$, $\Psi_{aaa}$, and additional
function  $\Phi$ [see Eqs.~(\ref{bin90}) and~(\ref{bin100})], all
of them evaluated at the critical point $a=a_c$, $\Lambda=
\Lambda_c$. One can easily show that each of these functions is a
solution of the linear problem
\begin{equation}
w^{\prime\prime}(x)-\Lambda_c\, Q_\psi\left(\Psi_c\right)\,
w(x)=S(x) \label{fo10}
\end{equation}
with different source terms $S(x)$ and different initial
conditions at $x=0$. The source terms and initial conditions are
listed in Table.~\ref{table}.

Let us show, as an example, the derivation of $S(x)$ and of the
initial conditions for two of the functions: $\Psi_a$ and
$\Psi_{\Lambda a}$. The starting point is Eq.~(\ref{cr10}) with
the initial conditions~(\ref{cr20}) and~(\ref{cr30}) for the
stripe solution $\Psi(x,a,\Lambda)$. The stripe solution depends
on $a$ and $\Lambda$ because they enter either the equation, or
the initial conditions. Let us differentiate the both sides of
Eq.~(\ref{cr10}) with respect to $a$. We obtain
\begin{equation}
\Psi_{a}^{\prime\prime}(x,a,\Lambda)- \Lambda\,
\frac{dQ}{d\psi}\Bigl|_{\psi=\Psi(x,a,\Lambda)}\,
\Psi_a(x,a,\Lambda) = 0\,. \label{fo100}
\end{equation}
Once $\Psi(x,a,\Lambda)$ is known, Eq. (\ref{fo100}) is a linear
differential equation for $\Psi_a$. Now let us differentiate, with
respect to $a$, Eqs. (\ref{cr20}) and~(\ref{cr30}). We obtain the
relations
\begin{equation}
\Psi_{a}(0,a,\Lambda)=1\,\,\, \mbox{and}\,\,\,
\Psi_{a}^{\prime}(0,a,\Lambda)=0\,, \label{fo110} \end{equation}
which serve as the initial conditions for the same function
$\Psi_a$. As we see, the source term $S(x)$ vanishes in this case.

Differentiating Eqs. (\ref{fo100}) and (\ref{fo110}) with respect
to $\Lambda$, we arrive at the following problem for the function
$\Psi_{a\Lambda}$:
$$
\Psi_{a\Lambda}^{\prime\prime}(x,a,\Lambda)- \Lambda\,
\frac{dQ}{d\psi}\Bigl|_{\psi=\Psi(x,a,\Lambda)}\,
\Psi_{a\Lambda}(x,a,\Lambda)=
$$
$$
\frac{dQ}{d\psi}\Bigl|_{\psi=\Psi(x,a,\Lambda)}\,
\Psi_{a}(x,a,\Lambda)+
$$
\begin{equation}
\Lambda\, \frac{d^2Q}{d\psi^2}\Bigl|_{\psi=\Psi(x,a,\Lambda)}\,
\Psi_{\Lambda}(x,a,\Lambda)\, \Psi_{a}(x,a,\Lambda) \, ,
\label{fo130}
\end{equation}
\begin{equation}
\Psi_{a\Lambda}(0,a,\Lambda)=0\, , \label{fo140}
\end{equation}
\begin{equation}
\Psi_{a\Lambda}^{\prime}(0,a,\Lambda)=0\, . \label{fo150}
\end{equation}
Again, once $\Psi, \Psi_a$ and $\Psi_{\Lambda}$ are known, Eq.
(\ref{fo130}) is a linear equation for $\Psi_{a \Lambda}$. Here
the source term is nonzero, while the initial
conditions~(\ref{fo140}) and (\ref{fo150}) are zero. Substituting
in Eqs.~(\ref{fo100})-(\ref{fo150}) $a=a_c$ and
$\Lambda=\Lambda_c$, we obtain the second and fourth rows of
Table~\ref{table}. The other rows of Table 1 can be obtained in a
similar way.

The functions $\Psi^\prime(x,a_c,\Lambda_c)$ and
$\Psi_a(x,a_c,\Lambda_c)$ are special, as they satisfy the
homogeneous form of Eq. (\ref{fo10}):  $S(x)=0$. Therefore, once
they are found, the rest of the functions from Table 1 can be
expressed through them:
$$
w(x)=\frac{1}{\Lambda_c\, Q(a_c)}\,
\Biggl[\Psi^{\prime}(x)\int\limits_0^x\Psi_a(\xi)\, S(\xi)\, d\xi
-
$$
\begin{equation}
-\Psi_a(x)\int\limits_0^x\Psi^{\prime}(\xi)\, S(\xi)\, d\xi\Biggr]
+C_1\, \Psi_a(x)+C_2\,\Psi^{\prime}(x)\, , \label{fo120}
\end{equation}
where $C_1$ and $C_2$ are integration constants. To satisfy the
zero initial conditions at $x=0$,  we must choose $C_1=C_2=0$ in
all cases. Furthermore, evaluating $w(x)$ at the thermal wall
$x=1$, we observe that the term proportional to $\Psi_a(x)$ in Eq.
(\ref{fo120}) vanishes, as $\Psi_a(1)=0$ at the critical point.
Therefore, for all functions from Table 1, except $\Psi^{\prime}$
and $\Psi_a$, we obtain
\begin{equation}\label{edge}
w(1,a_c,\Lambda_c)=\frac{\Psi^{\prime}(1)}{\Lambda_c\,Q(a_c)}\int\limits_0^1\Psi_a(\xi)\,
S(\xi)\, d\xi\,.
\end{equation}

\end{document}